
\catcode`@=11

\font\fourteenrm=cmr10 scaled\magstep2
\font\twelverm=cmr10 scaled\magstep1
\font\ninerm=cmr9            \font\sixrm=cmr6

\font\fourteenbf=cmbx10 scaled\magstep2
\font\twelvebf=cmbx10 scaled\magstep1
\font\ninebf=cmbx9            \font\sixbf=cmbx6
\font\seventeeni=cmmi10 scaled\magstep3     \skewchar\seventeeni='177
\font\fourteeni=cmmi10 scaled\magstep2      \skewchar\fourteeni='177
\font\twelvei=cmmi10 scaled\magstep1        \skewchar\twelvei='177
\font\ninei=cmmi9                           \skewchar\ninei='177
\font\sixi=cmmi6                            \skewchar\sixi='177
\font\seventeensy=cmsy10 scaled\magstep3    \skewchar\seventeensy='60
\font\fourteensy=cmsy10 scaled\magstep2     \skewchar\fourteensy='60
\font\twelvesy=cmsy10 scaled\magstep1       \skewchar\twelvesy='60
\font\ninesy=cmsy9                          \skewchar\ninesy='60
\font\sixsy=cmsy6                           \skewchar\sixsy='60

\font\fourteenex=cmex10 scaled\magstep2
\font\twelveex=cmex10 scaled\magstep1

\font\fourteensl=cmsl10 scaled\magstep2
\font\twelvesl=cmsl10 scaled\magstep1
\font\ninesl=cmsl9

\font\fourteenit=cmti10 scaled\magstep2
\font\twelveit=cmti10 scaled\magstep1
\font\twelvett=cmtt10 scaled\magstep1
\font\twelvecp=cmcsc10 scaled\magstep1
\font\tencp=cmcsc10
\newfam\cpfam
%
%
\newcount\f@ntkey            \f@ntkey=0
\def\samef@nt{\relax \ifcase\f@ntkey \rm \or\oldstyle \or\or
         \or\it \or\sl \or\bf \or\tt \or\caps \fi }
\def\fourteenpoint{\relax
    \textfont0=\fourteenrm          \scriptfont0=\tenrm
    \scriptscriptfont0=\sevenrm
     \def\rm{\fam0 \fourteenrm \f@ntkey=0 }\relax
    \textfont1=\fourteeni           \scriptfont1=\teni
    \scriptscriptfont1=\seveni
     \def\oldstyle{\fam1 \fourteeni\f@ntkey=1 }\relax
    \textfont2=\fourteensy          \scriptfont2=\tensy
    \scriptscriptfont2=\sevensy
    \textfont3=\fourteenex     \scriptfont3=\fourteenex
    \scriptscriptfont3=\fourteenex
    \def\it{\fam\itfam \fourteenit\f@ntkey=4
}\textfont\itfam=\fourteenit
    \def\sl{\fam\slfam \fourteensl\f@ntkey=5
}\textfont\slfam=\fourteensl
    \scriptfont\slfam=\tensl
    \def\bf{\fam\bffam \fourteenbf\f@ntkey=6
}\textfont\bffam=\fourteenbf
    \scriptfont\bffam=\tenbf     \scriptscriptfont\bffam=\sevenbf
    \def\tt{\fam\ttfam \twelvett \f@ntkey=7
}\textfont\ttfam=\twelvett
    \h@big=11.9\p@{} \h@Big=16.1\p@{} \h@bigg=20.3\p@{}
\h@Bigg=24.5\p@{}
    \def\caps{\fam\cpfam \twelvecp \f@ntkey=8
}\textfont\cpfam=\twelvecp
    \setbox\strutbox=\hbox{\vrule height 12pt depth 5pt width\z@}
    \samef@nt}
\def\twelvepoint{\relax
    \textfont0=\twelverm          \scriptfont0=\ninerm
    \scriptscriptfont0=\sixrm
     \def\rm{\fam0 \twelverm \f@ntkey=0 }\relax
    \textfont1=\twelvei           \scriptfont1=\ninei
    \scriptscriptfont1=\sixi
     \def\oldstyle{\fam1 \twelvei\f@ntkey=1 }\relax
    \textfont2=\twelvesy          \scriptfont2=\ninesy
    \scriptscriptfont2=\sixsy
    \textfont3=\twelveex          \scriptfont3=\twelveex
    \scriptscriptfont3=\twelveex
    \def\it{\fam\itfam \twelveit \f@ntkey=4
}\textfont\itfam=\twelveit
    \def\sl{\fam\slfam \twelvesl \f@ntkey=5
}\textfont\slfam=\twelvesl
    \scriptfont\slfam=\ninesl
    \def\bf{\fam\bffam \twelvebf \f@ntkey=6
}\textfont\bffam=\twelvebf
    \scriptfont\bffam=\ninebf     \scriptscriptfont\bffam=\sixbf
    \def\tt{\fam\ttfam \twelvett \f@ntkey=7
}\textfont\ttfam=\twelvett
    \h@big=10.2\p@{}
    \h@Big=13.8\p@{}
    \h@bigg=17.4\p@{}
    \h@Bigg=21.0\p@{}
    \def\caps{\fam\cpfam \twelvecp \f@ntkey=8
}\textfont\cpfam=\twelvecp
    \setbox\strutbox=\hbox{\vrule height 10pt depth 4pt width\z@}
    \samef@nt}
\def\tenpoint{\relax
    \textfont0=\tenrm          \scriptfont0=\sevenrm
    \scriptscriptfont0=\fiverm
    \def\rm{\fam0 \tenrm \f@ntkey=0 }\relax
    \textfont1=\teni           \scriptfont1=\seveni
    \scriptscriptfont1=\fivei
    \def\oldstyle{\fam1 \teni \f@ntkey=1 }\relax
    \textfont2=\tensy          \scriptfont2=\sevensy
    \scriptscriptfont2=\fivesy
    \textfont3=\tenex          \scriptfont3=\tenex
    \scriptscriptfont3=\tenex
    \def\it{\fam\itfam \tenit \f@ntkey=4 }\textfont\itfam=\tenit
    \def\sl{\fam\slfam \tensl \f@ntkey=5 }\textfont\slfam=\tensl
    \def\bf{\fam\bffam \tenbf \f@ntkey=6 }\textfont\bffam=\tenbf
    \scriptfont\bffam=\sevenbf     \scriptscriptfont\bffam=\fivebf
    \def\tt{\fam\ttfam \tentt \f@ntkey=7 }\textfont\ttfam=\tentt
    \def\caps{\fam\cpfam \tencp \f@ntkey=8 }\textfont\cpfam=\tencp
    \setbox\strutbox=\hbox{\vrule height 8.5pt depth 3.5pt width\z@}
    \samef@nt}
%
%
%
%
\newdimen\h@big  \h@big=8.5\p@
\newdimen\h@Big  \h@Big=11.5\p@
\newdimen\h@bigg  \h@bigg=14.5\p@
\newdimen\h@Bigg  \h@Bigg=17.5\p@
\def\big#1{{\hbox{$\left#1\vbox to\h@big{}\right.\n@space$}}}
\def\Big#1{{\hbox{$\left#1\vbox to\h@Big{}\right.\n@space$}}}
\def\bigg#1{{\hbox{$\left#1\vbox to\h@bigg{}\right.\n@space$}}}
\def\Bigg#1{{\hbox{$\left#1\vbox to\h@Bigg{}\right.\n@space$}}}
%
%
%
\normalbaselineskip = 20pt plus 0.2pt minus 0.1pt
\normallineskip = 1.5pt plus 0.1pt minus 0.1pt
\normallineskiplimit = 1.5pt
\newskip\normaldisplayskip
\normaldisplayskip = 20pt plus 5pt minus 10pt
\newskip\normaldispshortskip
\normaldispshortskip = 6pt plus 5pt
\newskip\normalparskip
\normalparskip = 6pt plus 2pt minus 1pt
\newskip\skipregister
\skipregister = 5pt plus 2pt minus 1.5pt
\newif\ifsingl@    \newif\ifdoubl@
\newif\iftwelv@    \twelv@true
\def\singlespace{\singl@true\doubl@false\spaces@t}
\def\doublespace{\singl@false\doubl@true\spaces@t}
\def\normalspace{\singl@false\doubl@false\spaces@t}
\def\Tenpoint{\tenpoint\twelv@false\spaces@t}
\def\Twelvepoint{\twelvepoint\twelv@true\spaces@t}
\def\spaces@t{\relax%
 \iftwelv@ \ifsingl@\subspaces@t3:4;\else\subspaces@t1:1;\fi%
 \else \ifsingl@\subspaces@t3:5;\else\subspaces@t4:5;\fi \fi%
 \ifdoubl@ \multiply\baselineskip by 5%
 \divide\baselineskip by 4 \fi \unskip}
\def\subspaces@t#1:#2;{
      \baselineskip = \normalbaselineskip
      \multiply\baselineskip by #1 \divide\baselineskip by #2
      \lineskip = \normallineskip
      \multiply\lineskip by #1 \divide\lineskip by #2
      \lineskiplimit = \normallineskiplimit
      \multiply\lineskiplimit by #1 \divide\lineskiplimit by #2
      \parskip = \normalparskip
      \multiply\parskip by #1 \divide\parskip by #2
      \abovedisplayskip = \normaldisplayskip
      \multiply\abovedisplayskip by #1 \divide\abovedisplayskip by #2
      \belowdisplayskip = \abovedisplayskip
      \abovedisplayshortskip = \normaldispshortskip
      \multiply\abovedisplayshortskip by #1
        \divide\abovedisplayshortskip by #2
      \belowdisplayshortskip = \abovedisplayshortskip
      \advance\belowdisplayshortskip by \belowdisplayskip
      \divide\belowdisplayshortskip by 2
      \smallskipamount = \skipregister
      \multiply\smallskipamount by #1 \divide\smallskipamount by #2
      \medskipamount = \smallskipamount \multiply\medskipamount by 2
      \bigskipamount = \smallskipamount \multiply\bigskipamount by 4
}
\def\normalbaselines{ \baselineskip=\normalbaselineskip
   \lineskip=\normallineskip \lineskiplimit=\normallineskip
   \iftwelv@\else \multiply\baselineskip by 4 \divide\baselineskip by
5
     \multiply\lineskiplimit by 4 \divide\lineskiplimit by 5
     \multiply\lineskip by 4 \divide\lineskip by 5 \fi }
\Twelvepoint  
\interlinepenalty=50
\interfootnotelinepenalty=5000
\predisplaypenalty=9000
\postdisplaypenalty=500
\hfuzz=1pt
\vfuzz=0.2pt
%
%
%
\def\pagecontents{
   \ifvoid\topins\else\unvbox\topins\vskip\skip\topins\fi
   \dimen@ = \dp255 \unvbox255
   \ifvoid\footins\else\vskip\skip\footins\footrule\unvbox\footins\fi
   \ifr@ggedbottom \kern-\dimen@ \vfil \fi }
\def\makeheadline{\vbox to 0pt{ \skip@=\topskip
      \advance\skip@ by -12pt \advance\skip@ by -2\normalbaselineskip
      \vskip\skip@ \line{\vbox to 12pt{}\the\headline} \vss
      }\nointerlineskip}
\def\makefootline{\baselineskip = 1.5\normalbaselineskip
                 \line{\the\footline}}
\newif\iffrontpage
\newif\ifletterstyle
\newif\ifp@genum
\def\nopagenumbers{\p@genumfalse}
\def\pagenumbers{\p@genumtrue}
\pagenumbers
\newtoks\paperheadline
\newtoks\letterheadline
\newtoks\letterfrontheadline
\newtoks\lettermainheadline
\newtoks\paperfootline
\newtoks\letterfootline
\newtoks\date
\footline={\ifletterstyle\the\letterfootline\else\the\paperfootline\fi}
\paperfootline={\hss\iffrontpage\else\ifp@genum\tenrm\folio\hss\fi\fi
}
\letterfootline={\hfil}
\headline={\ifletterstyle\the\letterheadline\else\the\paperheadline\fi}
\paperheadline={\hfil}
\letterheadline{\iffrontpage\the\letterfrontheadline
     \else\the\lettermainheadline\fi}
\lettermainheadline={\rm\ifp@genum page \ \folio\fi\hfil\the\date}
\def\monthname{\relax\ifcase\month 0/\or January\or February\or
   March\or April\or May\or June\or July\or August\or September\or
   October\or November\or December\else\number\month/\fi}
\date={\monthname\ \number\day, \number\year}
\countdef\pagenumber=1  \pagenumber=1
\def\advancepageno{\global\advance\pageno by 1
   \ifnum\pagenumber<0 \global\advance\pagenumber by -1
    \else\global\advance\pagenumber by 1 \fi \global\frontpagefalse }
\def\folio{\ifnum\pagenumber<0 \romannumeral-\pagenumber
           \else \number\pagenumber \fi }
\def\footrule{\dimen@=\prevdepth\nointerlineskip
   \vbox to 0pt{\vskip -0.25\baselineskip \hrule width 0.35\hsize
\vss}
   \prevdepth=\dimen@ }
\newtoks\foottokens
\foottokens={\Tenpoint\singlespace}
\newdimen\footindent
\footindent=24pt
\def\vfootnote#1{\insert\footins\bgroup  \the\foottokens
   \interlinepenalty=\interfootnotelinepenalty \floatingpenalty=20000
   \splittopskip=\ht\strutbox \boxmaxdepth=\dp\strutbox
   \leftskip=\footindent \rightskip=\z@skip
   \parindent=0.5\footindent \parfillskip=0pt plus 1fil
   \spaceskip=\z@skip \xspaceskip=\z@skip
   \Textindent{$ #1 $}\footstrut\futurelet\next\fo@t}
\def\Textindent#1{\noindent\llap{#1\enspace}\ignorespaces}
\def\footnote#1{\attach{#1}\vfootnote{#1}}

\def\foot{\attach\footsymbolgen\vfootnote{\footsymbol}}
\let\footsymbol=\star
\newcount\lastf@@t           \lastf@@t=-1
\newcount\footsymbolcount    \footsymbolcount=0
\newif\ifPhysRev
\def\footsymbolgen{\relax \ifPhysRev \iffrontpage \NPsymbolgen\else
      \PRsymbolgen\fi \else \NPsymbolgen\fi
   \global\lastf@@t=\pageno \footsymbol }
\def\NPsymbolgen{\ifnum\footsymbolcount<0
\global\footsymbolcount=0\fi
   {\iffrontpage \else \advance\lastf@@t by 1 \fi
    \ifnum\lastf@@t<\pageno \global\footsymbolcount=0
     \else \global\advance\footsymbolcount by 1 \fi }
   \ifcase\footsymbolcount \fd@f\star\or \fd@f\dagger\or \fd@f\ast\or
    \fd@f\ddagger\or \fd@f\natural\or \fd@f\diamond\or
\fd@f\bullet\or
    \fd@f\nabla\else \fd@f\dagger\global\footsymbolcount=0 \fi }
\def\fd@f#1{\xdef\footsymbol{#1}}
\def\PRsymbolgen{\ifnum\footsymbolcount>0
\global\footsymbolcount=0\fi
      \global\advance\footsymbolcount by -1
      \xdef\footsymbol{\sharp\number-\footsymbolcount} }
\def\space@ver#1{\let\@sf=\empty \ifmmode #1\else \ifhmode
   \edef\@sf{\spacefactor=\the\spacefactor}\unskip${}#1$\relax\fi\fi}
\def\attach#1{\space@ver{\strut^{\mkern 2mu #1} }\@sf\ }
%
%
\def\smallsize{\relax
\font\eightrm=cmr8
\font\eightbf=cmbx8
\font\eighti=cmmi8
\font\eightsy=cmsy8
\font\eightsl=cmsl8
\font\eightit=cmti8
\font\eightt=cmtt8
\def\eightpoint{\relax
\textfont0=\eightrm  \scriptfont0=\sixrm
\scriptscriptfont0=\sixrm
\def\rm{\fam0 \eightrm \f@ntkey=0}\relax
\textfont1=\eighti  \scriptfont1=\sixi
\scriptscriptfont1=\sixi
\def\oldstyle{\fam1 \eighti \f@ntkey=1}\relax
\textfont2=\eightsy  \scriptfont2=\sixsy
\scriptscriptfont2=\sixsy
\textfont3=\tenex  \scriptfont3=\tenex
\scriptscriptfont3=\tenex
\def\it{\fam\itfam \eightit \f@ntkey=4 }\textfont\itfam=\eightit
\def\sl{\fam\slfam \eightsl \f@ntkey=5 }\textfont\slfam=\eightsl
\def\bf{\fam\bffam \eightbf \f@ntkey=6 }\textfont\bffam=\eightbf
\scriptfont\bffam=\sixbf   \scriptscriptfont\bffam=\sixbf
\def\tt{\fam\ttfam \eightt \f@ntkey=7 }
\def\caps{\fam\cpfam \tencp \f@ntkey=8 }\textfont\cpfam=\tencp
\setbox\strutbox=\hbox{\vrule height 7.35pt depth 3.02pt width\z@}
\samef@nt}
\def\Eightpoint{\eightpoint \relax
  \ifsingl@\subspaces@t2:5;\else\subspaces@t3:5;\fi
  \ifdoubl@ \multiply\baselineskip by 5
            \divide\baselineskip by 4\fi }
\parindent=16.67pt
\itemsize=25pt
\thinmuskip=2.5mu
\medmuskip=3.33mu plus 1.67mu minus 3.33mu
\thickmuskip=4.17mu plus 4.17mu
\def\thinspace{\kern .13889em }
\def\negthinspace{\kern-.13889em }
\def\enspace{\kern.416667em }

\def\enskip{\hskip.416667em\relax}
\def\quad{\hskip.83333em\relax}
\def\qquad{\hskip1.66667em\relax}
\def\crr{\cropen{8.3333pt}}
\foottokens={\Eightpoint\singlespace}
\def\papersize{\vsize=38.67pc\hsize=29.17pc\hoffset=3.44pc\voffset=3.
7pc
               \skip\footins=\bigskipamount}
\def\lettersize{\hsize=5.417in\vsize=7.08in\hoffset=0in\voffset=.834i
n
   \skip\footins=\smallskipamount \multiply\skip\footins by 3 }
\def\attach##1{\space@ver{\strut^{\mkern 1.6667mu ##1} }\@sf\ }
\def\PH@SR@V{\doubl@true\baselineskip=20.08pt plus .1667pt minus
.0833pt
             \parskip = 2.5pt plus 1.6667pt minus .8333pt }
\def\author##1{\vskip\frontpageskip\titlestyle{\tencp ##1}\nobreak}
\def\address##1{\par\kern 4.16667pt\titlestyle{\tenpoint\it ##1}}
\def\andaddress{\par\kern 4.16667pt \centerline{\sl and} \address}
\def\SLAC{\address{Stanford Linear Accelerator Center\break
      Stanford University, Stanford, California, 94305}}
\def\abstract{\vskip\frontpageskip\centerline{\twelverm ABSTRACT}
              \vskip\headskip }
\def\submit##1{\par\nobreak\vfil\nobreak\medskip
   \centerline{Submitted to \sl ##1}}
\def\doeack{\foot{Work supported by the Department of Energy,
      contract $\caps DE-AC03-76SF00515$.}}
\def\cases##1{\left\{\,\vcenter{\Tenpoint\m@th
    \ialign{$####\hfil$&\quad####\hfil\crcr##1\crcr}}\right.}
\def\matrix##1{\,\vcenter{\Tenpoint\m@th
    \ialign{\hfil$####$\hfil&&\quad\hfil$####$\hfil\crcr
      \mathstrut\crcr\noalign{\kern-\baselineskip}
     ##1\crcr\mathstrut\crcr\noalign{\kern-\baselineskip}}}\,}
\Tenpoint \paperstyle
}
%
%
%
\newcount\chapternumber      \chapternumber=0
\newcount\sectionnumber      \sectionnumber=0
\newcount\equanumber         \equanumber=0
\let\chapterlabel=0
\newtoks\chapterstyle        \chapterstyle={\Number}
\newskip\chapterskip         \chapterskip=\bigskipamount
\newskip\sectionskip         \sectionskip=\medskipamount
\newskip\headskip            \headskip=8pt plus 3pt minus 3pt
\newdimen\chapterminspace    \chapterminspace=15pc
\newdimen\sectionminspace    \sectionminspace=10pc
\newdimen\referenceminspace  \referenceminspace=25pc
\def\chapterreset{\global\advance\chapternumber by 1
   \ifnum\the\equanumber<0 \else\global\equanumber=0\fi
   \sectionnumber=0 \makel@bel}
\def\makel@bel{\xdef\chapterlabel{%
\the\chapterstyle{\the\chapternumber}.}}
\def\sectionlabel{\number\sectionnumber \quad }
\def\alphabetic#1{\count255='140 \advance\count255 by
#1\char\count255}
\def\Alphabetic#1{\count255='100 \advance\count255 by
#1\char\count255}
\def\Roman#1{\uppercase\expandafter{\romannumeral #1}}
\def\roman#1{\romannumeral #1}
\def\Number#1{\number #1}
\def\unnumberedchapters{\let\makel@bel=\relax
\let\chapterlabel=\relax
\let\sectionlabel=\relax \equanumber=-1 }
\def\titlestyle#1{\par\begingroup \interlinepenalty=9999
     \leftskip=0.02\hsize plus 0.23\hsize minus 0.02\hsize
     \rightskip=\leftskip \parfillskip=0pt
     \hyphenpenalty=9000 \exhyphenpenalty=9000
     \tolerance=9999 \pretolerance=9000
     \spaceskip=0.333em \xspaceskip=0.5em
     \iftwelv@\fourteenpoint\else\twelvepoint\fi
   \noindent #1\par\endgroup }
\def\spacecheck#1{\dimen@=\pagegoal\advance\dimen@ by -\pagetotal
   \ifdim\dimen@<#1 \ifdim\dimen@>0pt \vfil\break \fi\fi}
\def\chapter#1{\par \penalty-300 \vskip\chapterskip
   \spacecheck\chapterminspace
   \chapterreset \titlestyle{\chapterlabel \ #1}
   \nobreak\vskip\headskip \penalty 30000
   \wlog{\string\chapter\ \chapterlabel} }
\let\chap=\chapter
\def\section#1{\par \ifnum\the\lastpenalty=30000\else
   \penalty-200\vskip\sectionskip \spacecheck\sectionminspace\fi
   \wlog{\string\section\ \chapterlabel \the\sectionnumber}
   \global\advance\sectionnumber by 1  \noindent
   {\sectionfont \enspace\chapterlabel \sectionlabel #1}\par
   \nobreak\vskip\headskip \penalty 30000 }
\def\subsection#1{\par
   \ifnum\the\lastpenalty=30000\else \penalty-100\smallskip \fi
   \noindent\undertext{#1}\enspace \vadjust{\penalty5000}}

\def\undertext#1{\vtop{\hbox{#1}\kern 1pt \hrule}}
\def\APPENDIX#1#2{\par\penalty-300\vskip\chapterskip
   \spacecheck\chapterminspace \chapterreset \xdef\chapterlabel{#1}
   \titlestyle{APPENDIX #2} \nobreak\vskip\headskip \penalty 30000
   \wlog{\string\Appendix\ \chapterlabel} }
\def\Appendix#1{\APPENDIX{#1}{#1}}
\def\appendix{\APPENDIX{A}{}}
%
%
%
\def\eqname#1{\relax \ifnum\the\equanumber<0%
     \xdef#1{{\noexpand\rm(\number-\equanumber)}}%
     \global\advance\equanumber by -1%
    \else \global\advance\equanumber by 1%
      \xdef#1{{\noexpand\rm(\chapterlabel \number\equanumber)}} \fi}

\def\eqn#1{\eqno\eqname{#1}#1}

\def\eqinsert#1{\noalign{\dimen@=\prevdepth \nointerlineskip
   \setbox0=\hbox to\displaywidth{\hfil #1}
   \vbox to 0pt{\vss\hbox{$\!\box0\!$}\kern-0.5\baselineskip}
   \prevdepth=\dimen@}}
\def\eqnalign#1{\eqname{#1}#1}
%

%
%
\def\GENITEM#1;#2{\par \hangafter=0 \hangindent=#1
    \Textindent{$ #2 $}\ignorespaces}
\outer\def\newitem#1=#2;{\gdef#1{\GENITEM #2;}}
\newdimen\itemsize                \itemsize=30pt
\newitem\item=1\itemsize;
\newitem\sitem=1.75\itemsize;     
\newitem\ssitem=2.5\itemsize;     
\outer\def\newlist#1=#2&#3&#4;{\toks0={#2}\toks1={#3}%
   \count255=\escapechar \escapechar=-1
   \alloc@0\list\countdef\insc@unt\listcount     \listcount=0
   \edef#1{\par
      \countdef\listcount=\the\allocationnumber
      \advance\listcount by 1
      \hangafter=0 \hangindent=#4
      \Textindent{\the\toks0{\listcount}\the\toks1}}
   \expandafter\expandafter\expandafter
    \edef\c@t#1{begin}{\par
      \countdef\listcount=\the\allocationnumber \listcount=1
      \hangafter=0 \hangindent=#4
      \Textindent{\the\toks0{\listcount}\the\toks1}}
   \expandafter\expandafter\expandafter
    \edef\c@t#1{con}{\par \hangafter=0 \hangindent=#4 \noindent}
   \escapechar=\count255}
\def\c@t#1#2{\csname\string#1#2\endcsname}
\newlist\point=\Number&.&1.0\itemsize;
\newlist\subpoint=(\alphabetic&)&1.75\itemsize;
\newlist\subsubpoint=(\roman&)&2.5\itemsize;
%

%
%
%
\newcount\referencecount     \referencecount=0
\newif\ifreferenceopen       \newwrite\referencewrite
\newtoks\rw@toks
\def\NPrefmark#1{\attach{\scriptscriptstyle [ #1 ] }}
\let\PRrefmark=\attach
\def\refmark#1{\relax\ifPhysRev\PRrefmark{#1}\else\NPrefmark{#1}\fi}
\def\refend{\refmark{\number\referencecount}}
\newcount\lastrefsbegincount \lastrefsbegincount=0
\def\refsend{\refmark{\count255=\referencecount
   \advance\count255 by-\lastrefsbegincount
   \ifcase\count255 \number\referencecount
   \or \number\lastrefsbegincount,\number\referencecount
   \else \number\lastrefsbegincount-\number\referencecount \fi}}
\def\refch@ck{\chardef\rw@write=\referencewrite
   \ifreferenceopen \else \referenceopentrue
   \immediate\openout\referencewrite=referenc.texauxil \fi}
%
{\catcode`\^^M=\active 
  \gdef\obeyendofline{\catcode`\^^M\active \let^^M\ }}%
%
{\catcode`\^^M=\active 
  \gdef\ignoreendofline{\catcode`\^^M=5}}
{\obeyendofline\gdef\rw@start#1{\def\t@st{#1} \ifx\t@st\blankend%
\endgroup \@sf \relax \else \ifx\t@st\bl@nkend \endgroup \@sf \relax%
\else \rw@begin#1
\backtotext
\fi \fi } }
{\obeyendofline\gdef\rw@begin#1
{\def\n@xt{#1}\rw@toks={#1}\relax%
\rw@next}}
\def\blankend{}
{\obeylines\gdef\bl@nkend{
}}
\newif\iffirstrefline  \firstreflinetrue
\def\rwr@teswitch{\ifx\n@xt\blankend \let\n@xt=\rw@begin %
 \else\iffirstrefline \global\firstreflinefalse%
\immediate\write\rw@write{\noexpand\obeyendofline \the\rw@toks}%
\let\n@xt=\rw@begin%
      \else\ifx\n@xt\rw@@d \def\n@xt{\immediate\write\rw@write{%
        \noexpand\ignoreendofline}\endgroup \@sf}%
             \else \immediate\write\rw@write{\the\rw@toks}%
             \let\n@xt=\rw@begin\fi\fi \fi}
\def\rw@next{\rwr@teswitch\n@xt}
\def\rw@@d{\backtotext} \let\rw@end=\relax
\let\backtotext=\relax

\newdimen\refindent     \refindent=30pt
\def\refitem#1{\par \hangafter=0 \hangindent=\refindent
\Textindent{#1}}
\def\REFNUM#1{\space@ver{}\refch@ck \firstreflinetrue%
 \global\advance\referencecount by 1 \xdef#1{\the\referencecount}}
\def\refnum#1{\space@ver{}\refch@ck \firstreflinetrue%
 \global\advance\referencecount by 1
\xdef#1{\the\referencecount}\refend}

\def\REF#1{\REFNUM#1%
 \immediate\write\referencewrite{%
 \noexpand\refitem{#1.}}%
\begingroup\obeyendofline\rw@start}
\def\ref{\refnum\?%
 \immediate\write\referencewrite{\noexpand\refitem{\?.}}%
\begingroup\obeyendofline\rw@start}
\def\Ref#1{\refnum#1%
 \immediate\write\referencewrite{\noexpand\refitem{#1.}}%
\begingroup\obeyendofline\rw@start}
\def\REFS#1{\REFNUM#1\global\lastrefsbegincount=\referencecount
\immediate\write\referencewrite{\noexpand\refitem{#1.}}%
\begingroup\obeyendofline\rw@start}
\newcount\figurecount     \figurecount=0
\newif\iffigureopen       \newwrite\figurewrite
\def\figch@ck{\chardef\rw@write=\figurewrite \iffigureopen\else
   \immediate\openout\figurewrite=figures.texauxil
   \figureopentrue\fi}
\def\FIGNUM#1{\space@ver{}\figch@ck \firstreflinetrue%
 \global\advance\figurecount by 1 \xdef#1{\the\figurecount}}
\def\FIG#1{\FIGNUM#1
   \immediate\write\figurewrite{\noexpand\refitem{#1.}}%
   \begingroup\obeyendofline\rw@start}
\def\fig{\FIGNUM\? fig.~\?%
\immediate\write\figurewrite{\noexpand\refitem{\?.}}%
\begingroup\obeyendofline\rw@start}
\def\figure{\FIGNUM\? figure~\?
   \immediate\write\figurewrite{\noexpand\refitem{\?.}}%
   \begingroup\obeyendofline\rw@start}
\def\Fig{\FIGNUM\? Fig.~\?%
\immediate\write\figurewrite{\noexpand\refitem{\?.}}%
\begingroup\obeyendofline\rw@start}
\def\Figure{\FIGNUM\? Figure~\?%
\immediate\write\figurewrite{\noexpand\refitem{\?.}}%
\begingroup\obeyendofline\rw@start}
\newcount\tablecount     \tablecount=0
\newif\iftableopen       \newwrite\tablewrite
\def\tabch@ck{\chardef\rw@write=\tablewrite \iftableopen\else
   \immediate\openout\tablewrite=tables.texauxil
   \tableopentrue\fi}
\def\TABNUM#1{\space@ver{}\tabch@ck \firstreflinetrue%
 \global\advance\tablecount by 1 \xdef#1{\the\tablecount}}
\def\TABLE#1{\TABNUM#1
   \immediate\write\tablewrite{\noexpand\refitem{#1.}}%
   \begingroup\obeyendofline\rw@start}
\def\Table{\TABNUM\? Table~\?%
\immediate\write\tablewrite{\noexpand\refitem{\?.}}%
\begingroup\obeyendofline\rw@start}
%

%
%
%
\def\masterreset{\global\pagenumber=1 \global\chapternumber=0
   \ifnum\the\equanumber<0\else \global\equanumber=0\fi
   \global\sectionnumber=0
   \global\referencecount=0 \global\figurecount=0
\global\tablecount=0 }
\def\FRONTPAGE{\ifvoid255\else\vfill\penalty-2000\fi
      \masterreset\global\frontpagetrue
      \global\lastf@@t=0 \global\footsymbolcount=0}

\def\paperstyle{\letterstylefalse\normalspace\papersize}
\def\letterstyle{\letterstyletrue\singlespace\lettersize}
\def\papersize{\hsize=35pc\vsize=50pc\hoffset=1pc\voffset=6pc
               \skip\footins=\bigskipamount}
\def\lettersize{\hsize=6.5in\vsize=8.5in\hoffset=0in\voffset=1in
   \skip\footins=\smallskipamount \multiply\skip\footins by 3 }
\paperstyle   
%
%
%
\def\MEMO{\letterstyle\FRONTPAGE \letterfrontheadline={\hfil}
    \line{\quad\fourteenrm SLAC
MEMORANDUM\hfil\twelverm\the\date\quad}
    \medskip \memod@f}

\def\memit@m#1{\smallskip \hangafter=0 \hangindent=1in
      \Textindent{\caps #1}}
\def\memod@f{\xdef\to{\memit@m{To:}}\xdef\from{\memit@m{From:}}%
     \xdef\topic{\memit@m{Topic:}}\xdef\subject{\memit@m{Subject:}}%
     \xdef\rule{\bigskip\hrule height 1pt\bigskip}}
\memod@f
\newskip\lettertopfil
\lettertopfil = -3.0truecm plus 0.0cm minus 0.0cm
\newskip\letterbottomfil
\letterbottomfil = 0pt plus 2.3in minus 0pt
\newskip\spskip \setbox0\hbox{\ } \spskip=-1\wd0
\def\addressee#1{\medskip\rightline{\the\date\hskip 30pt}\bigskip
   \vskip\lettertopfil
   \ialign to\hsize{\strut ##\hfil\tabskip 0pt plus \hsize \cr
#1\crcr}
   \medskip\noindent\hskip\spskip}
\newskip\signatureskip       \signatureskip=40pt
\def\signed#1{\par \penalty 9000 \bigskip \dt@pfalse
  \everycr={\noalign{\ifdt@p\vskip\signatureskip\global\dt@pfalse\fi}
}
  \setbox0=\vbox{\singlespace \halign{\tabskip 0pt \strut ##\hfil\cr
   \noalign{\global\dt@ptrue}#1\crcr}}
  \line{\hskip 0.5\hsize minus 0.5\hsize \box0\hfil} \medskip }

\def\endletter{\ifnum\pagenumber=1 \vskip\letterbottomfil\supereject
\else \vfil\supereject \fi}
\newbox\letterb@x
\def\lettertext{\par\unvcopy\letterb@x\par}
\def\multiletter{\setbox\letterb@x=\vbox\bgroup
      \everypar{\vrule height 1\baselineskip depth 0pt width 0pt }
      \singlespace \topskip=\baselineskip }
\def\letterend{\par\egroup}
%
%
%
\newskip\frontpageskip
\newtoks\pubtype
\newtoks\Pubnum
\newtoks\pubnum
\newif\ifp@bblock  \p@bblocktrue
\def\PH@SR@V{\doubl@true \baselineskip=24.1pt plus 0.2pt minus 0.1pt
             \parskip= 3pt plus 2pt minus 1pt }
\def\PHYSREV{\paperstyle\PhysRevtrue\PH@SR@V}
\def\titlepage{\FRONTPAGE\paperstyle\ifPhysRev\PH@SR@V\fi
   \ifp@bblock\p@bblock\fi}
\def\nopubblock{\p@bblockfalse}
\def\endpage{\vfil\break}
\frontpageskip=1\medskipamount plus .5fil
\pubtype={\tensl Preliminary Version}
\Pubnum={$\caps SLAC - PUB - \the\pubnum $}
\pubnum={0000}
\def\p@bblock{\begingroup \tabskip=\hsize minus \hsize
   \baselineskip=1.5\ht\strutbox \topspace-2\baselineskip
   \halign to\hsize{\strut ##\hfil\tabskip=0pt\crcr
   \the\Pubnum\cr \the\date\cr \the\pubtype\cr}\endgroup}
\def\title#1{\vskip\frontpageskip \titlestyle{#1} \vskip\headskip }
\def\author#1{\vskip\frontpageskip\titlestyle{\twelvecp #1}\nobreak}
\def\andauthor{\vskip\frontpageskip\centerline{and}\author}

\def\address#1{\par\kern 5pt\titlestyle{\twelvepoint\it #1}}
\def\andaddress{\par\kern 5pt \centerline{\sl and} \address}
\def\SLAC{\address{Stanford Linear Accelerator Center\break
      Stanford University, Stanford, California, 94305}}
\def\abstract{\vskip\frontpageskip\centerline{\fourteenrm ABSTRACT}
              \vskip\headskip }
\def\submit#1{\par\nobreak\vfil\nobreak\medskip
   \centerline{Submitted to \sl #1}}
\def\doeack{\foot{Work supported by the Department of Energy,
      contract $\caps DE-AC03-76SF00515$.}}
%
%
%
\def\ie{\hbox{\it i.e.}}     
\def\eg{\hbox{\it e.g.}}     

\def\\{\relax\ifmmode\backslash\else$\backslash$\fi}
\def\globaleqnumbers{\relax\ifnum\the\equanumber<0%
\else\global\equanumber=-1\fi}

\def\journal#1&#2(#3){\unskip, \sl #1~\bf #2 \rm (19#3) }
\def\cropen#1{\crcr\noalign{\vskip #1}}
\def\crr{\cropen{10pt}}
\def\topspace{\hrule height 0pt depth 0pt \vskip}
\def\coeff#1#2{\textstyle{#1\over #2}}

\let\int=\intop         
\def\prop{\mathrel{{\mathchoice{\pr@p\scriptstyle}{\pr@p\scriptstyle}
{
                \pr@p\scriptscriptstyle}{\pr@p\scriptscriptstyle} }}}
\def\pr@p#1{\setbox0=\hbox{$\cal #1 \char'103$}
   \hbox{$\cal #1 \char'117$\kern-.4\wd0\box0}}
\def\lsim{\mathrel{\mathpalette\@versim<}}
\def\gsim{\mathrel{\mathpalette\@versim>}}
\def\@versim#1#2{\lower0.2ex\vbox{\baselineskip\z@skip\lineskip\z@ski
p
  \lineskiplimit\z@\ialign{$\m@th#1\hfil##\hfil$\crcr#2\crcr\sim\crcr
}}}
%
%
%
%
\let\sec@nt=\sec
\def\sec{\relax\ifmmode\let\n@xt=\sec@nt\else\let\n@xt\section\fi\n@x
t}
\def\obsolete#1{\message{Macro \string #1 is obsolete.}}
\def\firstsec#1{\obsolete\firstsec \section{#1}}
\def\firstsubsec#1{\obsolete\firstsubsec \subsection{#1}}
\def\thispage#1{\obsolete\thispage
\global\pagenumber=#1\frontpagefalse}
\def\thischapter#1{\obsolete\thischapter \global\chapternumber=#1}
\def\nextequation#1{\obsolete\nextequation \global\equanumber=#1
   \ifnum\the\equanumber>0 \global\advance\equanumber by 1 \fi}
\def\BOXITEM{\afterassigment\B@XITEM\setbox0=}
\def\B@XITEM{\par\hangindent\wd0 \noindent\box0 }
%

%
\catcode`@=12 
\message{ by V.K.}
%
%
%
%
%
%
%
%
%
%
\catcode`@=11

\font\seventeencp=cmcsc10 scaled\magstep3
%
%
%
\newif\ifVM
\newif\ifVMreferences
\VMtrue\VMreferencestrue

\newif\ifsectionskip \sectionskiptrue

\newbox\leftpage \newdimen\fullhsize \newdimen\hstitle
\newdimen\hsbody
\newif\ifreduce
\reducefalse
\def\almostshipout#1{\if L\lr \count2=1
      \global\setbox\leftpage=#1 \global\let\lr=R
  \else \count2=2
    \shipout\vbox{
      \hbox to\fullhsize{\box\leftpage\hfil#1}}  \global\let\lr=L\fi}
%
%
\def\sevenbf{\sixbf}
%
%
\def\Smallsize{\smallsize\reducetrue
\let\lr=L
\hstitle=8truein\hsbody=4.75truein\fullhsize=24.3truecm\hsize=\hsbody
\output={
  \almostshipout{\leftline{\vbox{\makeheadline
  \pagebody\makefootline}}}\advancepageno
     }
\def\papersize{\SIZE\OFFSET\skip\footins=\bigskipamount}
\def\SIZE{\hsize=11.7truecm\vsize=17.5truecm}
\def\OFFSET{
\ifVM
\voffset=-1.4truecm\hoffset=-.312truecm
\else
\voffset=1.0truecm\hoffset=4.64truecm\fi}
\def\makeheadline{
\iffrontpage\line{\the\headline}
             \else\vskip .0truecm\line{\the\headline}\vskip .5truecm
\fi}
\def\makefootline{\iffrontpage\vskip  0.truecm\line{\the\footline}
               \vskip -.15truecm\line{\the\date\hfil}
              \else\line{\the\footline}\fi}
\paperheadline={
\iffrontpage\hfil
               \else
               \tenrm\hss $-$\ \folio\ $-$\hss\fi    }
\h@big=6.0\p@
\h@Big=8.5\p@
\h@bigg=11.5\p@
\h@Bigg=14.5\p@}
\def\SIZE{\hsize=16.73truecm\vsize=23.11truecm}
\def\OFFSET{
\ifVM
\voffset=0.4truecm\hoffset=-0.38truecm
\else
\voffset=3.05truecm\hoffset=1.8truecm\fi}
\def\papersize{\SIZE\OFFSET\skip\footins=\bigskipamount
\normaldisplayskip= 30pt plus 5pt minus 10pt}
\Pubnum={\rm CALT-68-\the\pubnum }
\def\title#1{\vskip\frontpageskip\vskip .50truein
     \titlestyle{\seventeencp #1} \vskip\headskip\vskip\frontpageskip
     \vskip .2truein}
\def\author#1{\vskip .27truein\titlestyle{#1}\nobreak}
\def\andauthor{\vskip .27truein\centerline{and}\author}
\def\p@bblock{\begingroup \tabskip=\hsize minus \hsize
   \baselineskip=1.5\ht\strutbox \topspace-2\baselineskip
   \halign to\hsize{\strut ##\hfil\tabskip=0pt\crcr
   \the \Pubnum\cr}\endgroup}
\def\makefootline{\iffrontpage\vskip .27truein\line{\the\footline}
                 \vskip -.1truein\line{\the\date\hfil}
              \else\line{\the\footline}\fi}
\paperfootline={\iffrontpage
 $\rm Ref.$\the\Pubnum\hfil\else\hfil\fi}
\paperheadline={
\iffrontpage\hfil
               \else
               \twelverm\hss $-$\ \folio\ $-$\hss\fi}
\newif\ifmref  
\newif\iffref  
\def\xrefsend{\xrefmark{\count255=\referencecount
\advance\count255 by-\lastrefsbegincount
\ifcase\count255 \number\referencecount
\or \number\lastrefsbegincount,\number\referencecount
\else \number\lastrefsbegincount-\number\referencecount \fi}}
\def\xrefsdub{\xrefmark{\count255=\referencecount
\advance\count255 by-\lastrefsbegincount
\ifcase\count255 \number\referencecount
\or \number\lastrefsbegincount,\number\referencecount
\else \number\lastrefsbegincount,\number\referencecount \fi}}
\def\xREFNUM#1{\space@ver{}\refch@ck\firstreflinetrue%
\global\advance\referencecount by 1
\xdef#1{\xrefend}}
\def\xrefend{\xrefmark{\number\referencecount}}
\def\xrefmark#1{[{#1}]}
\def\xRef#1{\xREFNUM#1\immediate\write\referencewrite%
{\noexpand\refitem{#1.}}\begingroup\obeyendofline\rw@start}%
\def\xREFS#1{\xREFNUM#1\global\lastrefsbegincount=\referencecount%
\immediate\write\referencewrite{\noexpand\refitem{#1.}}%
\begingroup\obeyendofline\rw@start}
\def\rrr#1#2{\relax\ifmref{\iffref\xREFS#1{#2}%
\else\xRef#1{#2}\fi}\else\xRef#1{#2}\xrefend\fi}
\def\multref#1#2{\mreftrue\freftrue{#1}%
\freffalse{#2}\mreffalse\xrefsend}
\referencecount=0
%

%
\space@ver{}\refch@ck\firstreflinetrue%
\immediate\write\referencewrite{}%
\begingroup\obeyendofline\rw@start{}%

\def\nup#1({Nucl.\ Phys.\ $\underline {B#1}$\ (}
\def\plt#1({Phys.\ Lett.\ $\underline  {#1}$\ (}
\def\cmp#1({Comm.\ Math.\ Phys.\ $\underline  {#1}$\ (}
\def\prp#1({Phys.\ Rep.\ $\underline  {#1}$\ (}
\def\prl#1({Phys.\ Rev.\ Lett.\ $\underline  {#1}$\ (}
\def\prv#1({Phys.\ Rev. $\underline  {#1}$\ (}
\def\und#1({            $\underline  {#1}$\ (}
\catcode`@=12
%
\def\coeff#1#2{{\textstyle { #1 \over #2}}\displaystyle}
\def\sectionfont{\bf}

\def\OFFSET{\hoffset=1.1truein\voffset=1.4truein}
\def\SIZE{\hsize=6truein}
\def\titlepage{\OFFSET\SIZE\frontpagetrue\paperstyle
{\hbox to6.0truein{\tenpoint \baselineskip=12pt
        \hfil\vtop{
       \hbox{\strut LBL-31104;UCB-PTH-91/39}
       \hbox{\strut USC-91/022}
 	\hbox{\strut CALT-68-1738}
        \hbox{\strut DOE RESEARCH AND}
        \hbox{\strut DEVELOPMENT REPORT}}}} }
\def\CALT{\smallskip
        \address{California Institute of Technology,
Pasadena, CA 91125}}
\def\author#1{\vskip .5in \centerline{#1}}
\def\andauthor#1{\smallskip \centerline{\it and} \smallskip
\centerline{#1}
}
\def\abstract{\vskip .5in \vfil \centerline{\twelvepoint \bf
Abstract}}
\def\support#1;{\foot{Work supported by #1.}}

\def\title#1{
\vskip .7truecm
\titlestyle{\fourteenrm #1}}
\def\makefootline{\iffrontpage\vskip
.27truein\line{\the\footline}
                 \vskip -.1truein\line{July, 1991\hfil}
              \else\line{\the\footline}\fi}
\paperfootline={\hfil}
\def\refitem#1{\par
\hangafter=0 \hangindent=\refindent \Textindent{#1}}
%
\def\LVW{\rrr\LVW{W.\ Lerche, C.\ Vafa and N.P.\
Warner,\nup324(1989) 427.}}
\def\LGREFS{\rrr\LGREFS{
C.\ Vafa and N.P.\ Warner, \plt218B  (1989) 51;
E. Martinec, \plt 217B(1989) 431;
D.\ Gepner, \plt 222B(1989) 207;
W.\ Lerche, C.\ Vafa and N.P.\ Warner,\nup324(1989) 427;
E. Martinec, {\it Criticality, catastrophes and
compactifications,}
V.G. Knizhnik memorial volume;
P.~Howe and P.\ West, \plt223B(1989) 377;
K. Ito, \plt231B(1989) 125;
S.\ Cecotti, L.\ Girardello and A.\ Pasquinucci,
\nup328(1989)701,
Int.\ J.\ Mod.\ Phys.\ A6 (1991) 2427;
S.\ Cecotti and L.\ Girardello, \nup328(1989) 701;
S.\ Cecotti, Int.\ J,\ Mod.\ Phys.\ A6 (1991) 1749,
\nup355(1991) 755.}}
\def\FRKL{\rrr\FRKL{See for example,
F. Klein, {\it Vorlesungen \"uber die
Theorie der elliptischen Modulfunktionen}, Teubner, Leipzig
1890.}}
\def\Arn{\rrr\Arn{V.I. Arnold,
{\it Singularity Theory}, Lond. Math.
Soc. Lecture Series \#53; V.I. Arnold, S.M. Gusein-Zade and
A.N. Varchenko, {\it Singularities of differentiable maps,}
Birkh\"auser (1985).}}
\def\KS {\rrr\KS  {Y.\ Kazama and H.\ Suzuki, \plt216B(1989)
112; \nup321(1989) 232.}}
\def\EYa{\rrr\EYa{T. Eguchi and
 S-K. Yang, Phys. Lett. B 224 (1989) 373.}}
\def\Strom {\rrr\Strom  {A.\ Strominger, \cmp133 (1990)
163.}}
\def\BVar {\rrr\BVar  {B. Blok and  A.\ Varchenko, {\it
Topological
 conformal field theories and the flat coordinates},
preprint
 IASSNS-HEP-91/5.}}
\def\CeVa {\rrr\CeVa  {S.\ Cecotti and C.\ Vafa, {\it
Topological Anti-Topological Fusion}, preprint HUTP-91/A031
and
SISSA-69/91/EP.}}
\def\Bott {\rrr\Bott  {M.F.\ Atiyah, R.\ Bott anf L.\
G\"aarding, Acta Mathematica 131 (1973) 145.}}
\def\Griff {\rrr\Griff  {P.\ Griffiths, Ann. Math. 90 (1969)
460.}}
\def\CandA {\rrr\CandA  {P.\ Candelas, \nup 298 (1988)
458.}}
\def\EVNW {\rrr\EVNW  {E.\ Verlinde and N.P.\ Warner, {\it
Topological
Landau-Ginzburg Matter at c=3}, preprint USC-91/005,
IASSNS-HEP-91/16, to appear in Phys. Lett. B.}}
\def\CandB {\rrr\CandB  {P.\ Candelas, X.C.\ de la Ossa,
P.S.\ Green and
L.\ Parkes, {\it  A Pair of Calabi-Yau Manifolds as an
Exactly Soluble Superconformal Theory}, U.\ of Texas
preprint UTTG-25-1990.}}
\def\LCrn{\rrr\LCrn  {L.\ Crane, Commun. Math. Phys. 110
(1987) 391.}}
\def\GDB{\rrr\GDB  {G.D. Birkhoff, Math. Ann. 74 (1913)
122.}}
\def\ZMa{\rrr\ZMa  {Z.\ Maassarani, {\it On the Solution of
Topological
Landau-Ginzburg Models with c=3},  USC preprint USC-
91/023.}}
\def\KUh{\rrr\KUh {K.\ Uhlenbeck, private communication.}}
\def\Wit{\rrr\Wit{B.\ de Wit and A.\ van Proeyen, \nup245
(1984) 89;
E.\ Cremmer, C.\ Kounnas, A. \ van Proeyen, J.P.\
Derendinger,
S.\ Ferrara, B.\ de Wit and L.\ Girardello,  \nup250 (1985)
385;
S.\ Cecotti, S.\ Ferrara and L.\ Girardello,
Int.\ Mod.\ J.\ Phys.\ A4 (1989) 2475.}}
\def\Flatco{\rrr\Flatco{S.\ Cecotti, \cmp131(1990) 517,
\cmp124(1989) 23; S.\ Ferrara and A.\ Strominger, {\it N=2
spacetime
supersymmtry and Calabi-Yau moduli space}, preprint CERN-
TH-5291/89;
P.\ Candelas
and X.C.\ de la Ossa, \nup355 (1991) 455;
A.\ Stromiger, \cmp133 (1990) 163;
L.J.\ Dixon, V.S.\ Kaplunovsky and J.\ Louis,  \nup329
(1990) 27;
L.\ Castellani, R.\ D'Auria and S.\ Ferrara, \plt241B (1990)
57,
Class. Quantum Grav. 7 (1990) 1767.}}
\def\Katz {\rrr\Katz {N.\ Katz, Publ.\ Math. I.H.E.S. 35
(1968) 71.}}
\def\GiSm {\rrr\GiSm  {A.\ Giveon and  D.-J.\ Smit, Mod.
Phys. Lett. A {\bf 6} No. 24 (1991) 2211.}}
\def\Vafa {\rrr\Vafa  {C.\ Vafa, Mod.\ Phys.\ Lett. A6
(1991) 337.}}
\def\DVV {\rrr\DVV {R.\ Dijkgraaf, E. Verlinde and H.
Verlinde,
\nup352(1991) 59.}}
\def\Ferr {\rrr\Ferr  {A.\ Cadavid and S.\ Ferrara, {\it
Picard-Fuchs equations and the moduli space of
superconformal field theories}, preprint
CERN-Th.-6154.}}
\def\LLW {\rrr\LLW  {W.\ Lerche, D.\ L\"ust and N.P.\
Warner,
\plt231 (1989) 417.}}
\def\CandC {\rrr\CandC  {P.\ Candelas and X.\ de la Ossa,
\nup355 (1991) 455.}}
\def\Saito {\rrr\Saito  {K.\ Saito, Publ. RIMS, Kyoto Univ.,
19 (1983) 1231.}}
\def\FaT {\rrr\Fat  {L.D.\ Faddeev and L.A.\ Takhtajan, {\it
Hamiltonian Methods in the Theory of Solitons}, Springer
(1989).}}
\def\Dirk {\rrr\Dirk  {A.\ Giveon and  D.-J.\ Smit, {\it
Properties of superstring vacua from
(topological) Lan\-dau-Ginz\-burg models}, Berkeley preprint
LBL-30831, June 1991.}}
\def\Brieskorn {\rrr\Brieskorn  {E.\ Brieskorn, Nova Acta
Leopoldina, NF, {\bf 240} (1981) 65.}}
\def\yano {\rrr\yano  {K. Saito, T. Yano, J. Sekiguchi,
Commun.\ in Algebra, {\bf 8} (1980), 373.}}
\def\kato {\rrr\kato  {M. Kato and S. Watanabe, Bull. Coll.
Sci., Univ Ryukus, {\bf 32} (1981) 1.}}
\def\Forsyth {\rrr\Forsyth  {A. Forsyth, {\it Theory of
Differential Equations}, Vol. 4
pp. 174, Dover Publications, New-York (1959).}}
\def\AGDJS {\rrr\AGDJS  {A. Giveon and D.-J. Smit,
\nup349(1991) 168.}}
\def\Jan {\rrr\Jan  {J.\ Louis, private communication.}}
\def\Nikulin {\rrr\Nikulin  {I. Piatecki-Shapiro and I.
Shafarevich, Math. USSR Izvestia, {\bf 5} (1971) 547.}}
\def\slat {\rrr\slat  {L.\ Slater, {\it Generalized
hypergeometric functions},
Cambridge University Press (1966).}}
\def\DIZ {\rrr\DIZ  {P.\ Di Francesco, C.\ Itzykson and J.-
B.\ Zuber, {\it Classical $W$-Algebras}, Princeton preprint
PUPT-1211 (1990).}}
\def\Morr {\rrr\Morr  {D. Morrison, {\it Mirror Symmetry and
Rational Curves in Quintic Threefolds: A Guide for
Mathematicians}, Duke preprint DUK-M-91-01 (1991).}}
%

\def\Coe#1.#2.{{#1\over #1}}
\def\coeff#1#2{\relax{\textstyle {#1 \over
#2}}\displaystyle} 
\def\coe#1.#2.{\relax{\textstyle {#1 \over
#2}}\displaystyle} 
\def\half{{1 \over 2}}
\def\shalf{\relax{\textstyle {1 \over 2}}\displaystyle} %

\def\to{\rightarrow}
\def\notin{\hbox{{$\in$}\kern-.51em\hbox{/}}}

\let\bar=\overline

\def\sumpr#1{\mathop{{\sum}'}}

\def\attac#1{\Bigl\vert
{\phantom{X}\atop{{\rm\scriptstyle #1}}\phantom{X}}}
\def\del{\partial}

\def\inbar{\vrule height1.5ex width.4pt depth0pt}
\def\IB{\relax{\rm I\kern-.18em B}}
\def\IC{\relax\,\hbox{$\inbar\kern-.3em{\rm C}$}}
\def\ID{\relax{\rm I\kern-.18em D}}
\def\IE{\relax{\rm I\kern-.18em E}}
\def\IF{\relax{\rm I\kern-.18em F}}
\def\IG{\relax\,\hbox{$\inbar\kern-.3em{\rm G}$}}
\def\IH{\relax{\rm I\kern-.18em H}}
\def\II{\relax{\rm I\kern-.18em I}}
\def\IK{\relax{\rm I\kern-.18em K}}
\def\IL{\relax{\rm I\kern-.18em L}}
\def\IM{\relax{\rm I\kern-.18em M}}
\def\IN{\relax{\rm I\kern-.18em N}}
\def\IO{\relax\,\hbox{$\inbar\kern-.3em{\rm O}$}}
\def\IP{\relax{\rm I\kern-.18em P}}
\def\IQ{\relax\,\hbox{$\inbar\kern-.3em{\rm Q}$}}
\def\IR{\relax{\rm I\kern-.18em R}}
\font\sanse=cmss12
\def\ZZ{\relax{\hbox{\sanse Z\kern-.42em Z}}}
\def\a{\alpha} \def\b{\beta}   

\def\G{\Gamma} \def\l{\lambda} 
\def\s{\sigma}
\def\cA{{\cal A}} \def\cB{{\cal B}} 
\def\cD{{\cal D}}
\def\cF{{\cal F}} \def\cG{{\cal G}}

 \def\cO{{\cal O}} 

\def\cR{{\cal R}} 

\def\nex#1{$N\!=\!#1$}                          
\def\OFFSET{\voffset=0.35truein\hoffset=0.25truein}
\def\SIZE{\hsize=6truein}
\def\LG{Lan\-dau-Ginz\-burg}

\def\Gminus{G^-_{-\half}}
\def\ie{{\it i.e.}}
\def\Om{\Omega}
\def\om{\omega}
\def\vp{\varpi}
\def\a{\alpha}
\def\be{\beta}
\def\at{\a(t)}
\def\b#1{\beta_{#1}}
\def\mol#1.{(-1)^{#1}}
\def\l{\lambda}
\def\lp#1{{\l+#1}}
\def\G#1.{\Gamma(#1)}
\def\w#1.#2.{{#1\over W^{#2}}}
\def\dx{dx^1\!\!\wedge\! dx^2\!\dots\!\wedge\! dx^n}
\def\intw#1{\int\!\dx\w1.\lp#1.}
\def\oat{(1-\a^3)}
\def\oam{(1-\a^N)}
\def\sBig#1{{\phantom{\Big#1}}\!\!\!}
\def\xa{x^A}
\def\s{s}
\def\si{{\s_i}}
\def\sp{\s'}
\def\so{{\s_1}}
\def\pa{\phi_\a(\xa;\sp)}
\def\u#1#2.{u_{#1}^{(#2)}}
\def\Bia#1.{\cB_{i\ \a}^{(#1)\ \be}(\sp)}
\def\Aa#1.{\cA_\a^{(#1)\ \be}(\sp)}
\def\ddx{{\del\over\del\xa}}
\def\cp{\IC\!P^{d+1}}
\def\nn{^{(N)}}
\def\bi#1.#2.{b_{#2}^{(#1)}}
\def\dada#1.{{\del^{#1}\over\del\a^{#1}}}
\def\frac#1#2{{#1\over #2}}
\pubnum={xxxx}
\pubtype={}
\titlepage
\title{Differential Equations for Periods and Flat
Coordinates \break
in Two Dimensionsional Topological Matter Theories}
\vskip-.8truecm
\author{\ W.$\,$Lerche
\foot{Address after Nov.\ 1, 1991: CERN, Geneva,
Switzerland.}
}
\vskip-.3truecm
\CALT
\andauthor{D.-J. Smit }
\vskip-.3truecm
\address{Department of Physics, University of California,
Berkeley, CA 94720}
\andauthor{\ N.P.\ Warner}
\vskip-.3truecm
\address{Physics Department, U.S.C., University Park,
Los Angeles, CA 90089}
\vskip -1.truecm
\abstract
\penalty1000
{We consider two dimensional topological \LG\ models.  In
order to obtain the
free energy of these models, and to determine the K\"ahler
potential for the
marginal perturbations, one needs to determine flat or
`special' coordinates that
can be used to parametrize the perturbations of the
superpotentials.
This paper describes the relationship between the natural
\LG\ parametrization
and these flat coordinates.  In particular we show how one
can explicitly obtain
the differential equations that relate the two.  We discuss
the problem for both
Calabi-Yau manifolds and for general topological matter
models (with arbitary
central charges)
with relevant and marginal perturbations.  We also give a
number of examples.}
\penalty-1000 \endpage
\chapternumber=0
\pagenumber=1

\chap{Introduction}

Topological \LG\ theories\foot
{By this we mean the topological, chiral primary sub-sector
of
two-dimensional \nex2 supersymmetric \LG\ models, or
equivalently,
topologically twisted \LG\ models \Vafa.}
 are not only of interest in their own right, but they also
determine the modular
dependence of the Yukawa couplings in string theories \GiSm.
 The correlation functions of such topological models \DVV\
are completely
determined by a prepotential (or free energy), ${\cal F}$,
and in particular there is a set of `flat' (or `special')
\multref\Flatco{\Strom\BVar\CeVa}\
coordinates, $t_i$, $i =1, \dots, \mu$, in which the three
point function can be written as
$$
C_{ijk} ~=~ {\partial^3\ {\cal F} \over \partial t_i\partial
t_j\partial t_k} , \eqn\CFreln
$$
and for which
$$
{C_{ij}}^m ~ C_{klm} ~=~ {C_{il}}^m ~ C_{kjm} . \eqn\Ccons
$$
These coordinates are referred to as flat since the two
point function, $\eta_{ij}$,
is an invertible, $t$-independent matrix, providing a
natural, flat metric on the space of
chiral primary fields.  One set of flat coordinates is
provided by taking the $t_i$ to
be the coupling constants of the chiral primary
perturbations about the underlying
$N=2$ superconformal field theory.  In particular one has:
$$
C_{ijk}(t) ~\equiv ~ \bigg\langle~ \phi_i ~\phi_j ~\phi_k ~
exp\Big[ \sum_\ell t_\ell \int\! d^2z\, \phi^{(1,1)}_\ell
(z, \bar z) \Big] \bigg\rangle \eqn\Cdefn
$$
where $\phi^{(1,1)}_\ell\equiv\Gminus \widetilde \Gminus
\phi_\ell$.
One could, in principle, consider perturbations by any
chiral primary field;
however, for several reasons it is natural, and perhaps
necessary
\Vafa, to restrict ones attention to relevant and marginal
perturbations.

In string theory, only marginal perturbations are considered
(relevant operators
would generate space-time tachyons and thus are projected
out).  The
three-point functions, or structure constants, $C_{ijk}$,
determine the Yukawa
couplings of the low-energy effective field theory.
  In addition to this, the K\"ahler potential, $K$, of the
Zamolodchikov metric is
determined from the prepotential ${\cal F}$.  To obtain $K$
one first passes to
homogeneous coordinates $z^A$, $A= 0, \dots, \mu$, such that
$t_i = z^i/z^0$; $i
=1,\dots, \mu$ and views $\cF (t_i)$ as a function of the
$z^A$ that is
homogeneous of degree $2$.  That is, $\cF (z^A) \equiv
(z^0)^2 \cF (z^A/z^0) =
(z^0)^2 \cF (t_i)$.  One then has \Wit\Flatco\foot{
It appears \Jan\ that $K$ is not uniquely defined given flat
coordinates that obey only \CFreln. We believe that the
`correct' flat coordinates are those
which obey \Ccons\ as well.} :
$$
K = -log ~\bigg(~i~ \bar t^{\bar \jmath} {{\partial {\cal
F}} \over {\partial t^{j}}} -
{}~ i ~ t^{j}  {{\partial {\bar {\cal F}}}
\over {\partial \bar t^{\bar \jmath}}} ~\bigg). \eqn\KFreln
$$

In this paper we will consider topological (or \nex2
supersymmetric) theories that have a
\LG\ description \LGREFS.  One can obtain a set of
coordinates for such a
topological field theory simply by parametrizing the
superpotential, $W$ \DVV.
The problem is that these parameters are generally not the
{\it flat} coordinates.
One needs the flat coordinates to use \KFreln. Moreover, for
general coordinates
the derivatives in \CFreln\ are covariant, making it
difficult to determine ${\cal F}$
from $C_{ijk}$.
However, once given a parametrization of $W$ in terms of
flat coordinates $t_i$,
it is trivial to determine the structure constants
$C_{ijk}(t)$ via simple
polynomial multiplication modulo the vanishing relations:
$$
\phi_i(t)\,\phi_j(t)\ =\ {C_{ij}}^k(t) \, \phi_k(t) \quad
{\rm mod} \quad \nabla
W\equiv0 \ , \eqn\polymult
$$
where $\phi_i(t)\equiv - {\del\over\del t_i}W(t)$ \DVV.

  Thus our purpose will be to determine how these flat
coordinates can be related to general parametrizations of
the Landau-Ginzburg potential.

To date there have been several approaches to solving this
problem.  On Calabi-
Yau manifolds the required coordinates can be related to the
periods of the
holomorphic $3$-form evaluated on an integral homology basis
\multref\Strom{\CandB\CandC}.  One can sometimes evaluate
these periods explicitly as in
\CandB.  One also knows that such periods must satisfy a
linear differential
equation, and it turns out that there is an elementary
algorithm for determining
this differential equation directly from the \LG\
superpotential $W$. (A brief
exposition of this has already been given in \Dirk .)   This
method has been well
known to mathematicians for many years (see, for example,
\multref\Katz\Griff\Morr),
but is apparently not well known in the physics community,
and so
we will give an exposition of the procedure, along with some
examples, in section 2.

In section 3 we will discuss the relationship between flat
coordinates and the
differential equations of section 2, and derive in detail
the flat coordinate
of a family of $K3$ surfaces.   We will also discuss the
role of the duality group of the \LG\ potential.
 In section 4 we will describe how the Calabi-Yau techniques
can be generalized
to general topological \LG\ models.  This time one considers
periods of
differential forms on the level curves of $W$, and then one
shows that by choosing the
gauge carefully, one can solve the consistency conditions
\CFreln\ and \Ccons.
The basic method is also known in the mathematics literature
and is an
application of the work of K.~Saito.  A recent, rather brief
exposition of this
appeared in \BVar.  Our intention here is not only to
simplify the exposition still
further, but also to show that if one restricts to relevant
and marginal
perturbations then the calculations can be simplified.
Indeed
(contrary to the expectations expressed in \BVar) it becomes
relatively
straightforward to solve topological models whose underlying
conformal theory
has $c >3$.

\chap{Chiral Rings and Differential Equations for Periods}

In this section we will, for simplicity, consider a $d$-
dimensional
(non-singular) hypersurface, $V$, defined by the vanishing
of a homogenous
polynomial, $W$, of degree $\nu$ in $\cp$ (the
generalization
to weighted projective spaces is elementary). We will denote
the
homogenous coordinates on $\cp$ by $\xa, A=1,\dots,d+2$.
We will also consider $W$ to be a function of the $\xa$ and
of some
(dimensionless) moduli $\mu_i$.

 If the first Chern class of $V$
vanishes then there is a globally defined, holomorphic $d$-
form,
$\Om$, on $V$. This form can be represented
\multref\Bott{\Griff\CandA}\ by
$$
\Om\ =\ \int_\gamma\w1.{}.\,\om\ ;\qquad\ \
\omega=\sum_{A=1}^{d+2}(-
1)^Ax^A\,dx^1\!\wedge\!\dots\wedge\!{\widehat{dx^A}}
\!\wedge\!\dots\!\wedge\!dx^{d+2}\ ,  \eqn\omeg
$$
where $\gamma$ is a small, one-dimensional curve winding
around the
hypersurface $V$. More generally, the integral
$$
\Om_\a\ =\ \int_\gamma\w p_\a(\xa).k+1.\,\om\ ,\eqn\omegal
$$
where $p_\a(\xa)$ is a homogenous polynomial of degree
$k\nu$, represents a
(rational) differential $d$-form.  The form,
$\Omega_\alpha$, is an element of
$\bigoplus_{q=0}^k\bigwedge^{\!(d-q,q)}$. One finds \Griff
\CandA\ that
$\Om_\a$ represents a non-trivial cohomology element in
$F_k\equiv\bigoplus_{q=0}^k H^{(d-q,q)}$$(V,\IR)$ if and
only if
$p_\a$ is a non-trivial element of the local ring, $\cR$, of
$W$. If we take
the $p_\a$ to be a basis for $\cR$, then the corresponding
forms,
$\Om_\a$, are a basis for the cohomology $H^d$. For the
moment we will restrict
our attention
to these cohomologically non-trivial differential forms.

The set of periods of a differential form, $\Om_\a$, is
defined to be the
integrals of $\Om_\a$ over elements of a basis of the
integral homology of
$V$. This also has a convenient representation:
$$
\Pi_\a^{\ \be}\ = \int_{\Gamma_\be}\w p_\a(\xa).k+1.\,\om\
,\eqn\periodef
$$
where $\Gamma_\be$ is a representative of a homology basis
in
$H_{d+1}(\cp-V,\ZZ)$. The curve $\Gamma_\be$ may be thought
of as a tube over the corresponding cycle in $H_d(V,\ZZ)$.

{}From now on, we will fix $\Gamma_\be$ and consider
the vector $\vp_\a\equiv\Pi_\a^{\ \be}$. Considered as a
function of
the moduli $\mu_i$ of $W$, the vector $\vp$ satisfies a
regular singular,
matrix differential equation
$$
\left[{\del\over\del\mu_i}\ -\ A_i(\mu_j)\right]\,\vp\ =\
0\eqn\matdeq
$$
for some matrices $A_i$. (The complete set of solutions to
this
differential equation is in fact all of the columns of the
period matrix
$\Pi_\a^{\ \be}$ \Griff\Arn.)

Our purpose in this section is to give an elementary
procedure that
generates the differential equation directly from $W$. The
key ingredient
is a technical result established in \Griff. That is, one
considers
a differential $(d-1)$-form of $V$ defined by
$$
\phi\!=\!\!\int_\gamma\!\!\w1.l.\Big\{\!\sum_{B<C}(-
1)^{B+C}\big[x^BY_C(\xa)\!-
\!x^CY_B(\xa)\big]\!dx^1\!\wedge\dots\wedge\!{\widehat{dx^B}
}
\!\wedge\dots\wedge\!{\widehat{dx^C}}\!\wedge
\dots\wedge dx^{d+2}\!\Big\}\!,   \eqn\dmoneform
$$
where the $Y_B(\xa)$ are homogenous of degree $l\nu-(d+1)$.
One finds that
$$
d\phi\ =\
\int_\gamma\w1.l+1.\Big[\,l\,\big(\sum_{A=1}^{d+2}Y_A{\del
W\over
\del x^A}\big) - W\big(\sum_{A=1}^{d+2}{\del Y_A\over
\del x^A}\big)\Big]\om\ .\eqn\starstar
$$
Because this is an exact form, it provides us with a simple
means of
integrating by parts. Equivalently, if $p_\a(\xa)$ in
\omegal\ has the form
$\sum Y_A{\del W\over\del x^A}$ then, modulo exact forms, we
have from
\starstar
$$
\Om_a\ \equiv\ {1\over
k}\int_\gamma\Big[\w1.k.\sum_A\big({\del Y_A\over
\del x^A}\big)\Big]\,\om\ .  \eqn\reduced
$$
One can iterate this procedure (if necessary) and so reduce
the numerator until
it lies in the local ring of $W$. Note that this procedure
amounts to the most naive
form of partial integration.

To derive the differential equation, simply differentiate
under the integral
to obtain
$$
{\del\vp_\a\over \del\mu_i}\ =\
\int_\Gamma\Big[\w(\del_{\mu_i} p_\a).k+1. -
  (k+1)\w p_\a(\del_{\mu_i}W).k+2.\Big]\,\om\ , \eqn\wderiv
$$
and then partially integrate until all numerators have been
reduced to
elements of the local ring of $W$. Expressing this reduced
r.h.s.\
of \wderiv\ in terms of the $\Om_\a$ immediately yields
\matdeq.

If one is interested in the dependence of $\Om_\a$ on one
particular
modulus, $\mu_0$, then one can reduce the first order system
\matdeq\ to
one linear, regular singular O.D.E.\ for
$\vp_1\equiv\int\Om$
of order equal to or less than the dimension, $\mu$, of
the local ring of $W$. Note that the order is often much
less then $\mu$.
For example, if $p_\a(\xa)$ and $W(\xa;\mu_0)$ are invariant
under some
discrete symmetry, then the foregoing reduction procedure
can generate only
those $\Om_\be$ for which $p_\be$ is also invariant.  Hence
the order of the
differential equation cannot be greater than the number of
such invariant
$p_\beta$'s.

We conclude this section by calculating a couple of
examples. First we
consider the cubic torus, that is, we take $d=1$,
$(\mu_0\equiv\a)$, and:
$$
W(\xa)\ =\ \coeff13(x^3+y^3+z^3) - \a\,xyz\ \ .\eqn\torusex
$$
Let $\vp_1=\int_\Gamma\w1.{}.\,\om$ and $\vp_2=\int_\Gamma\w
xyz.2.\,
\om$. Then obviously ${\del\over\del\a}\vp_1=\vp_2$ and
$$
{\del\over\del\a}\vp_2\ =\ 2 \int\w x^2y^2z^2.3.\,\om\ .
$$
One now uses the identity:
$$
 \oat\, x^2y^2z^2=xz^2\Big\{
 \a^2 y\del_zW+\a z\del_xW+x\del_yW\Big\}\ ,\eqn\xyzexp
$$
and integrating by parts yields
$$
(1-\a^3){\del\over\del\a}\vp_2\ =\ \a\int\w z^3.2.\,\om +
2\a^2\int\w xyz.2.\,
\om\ .
$$
Using $z^3=z\del_zW+\a xyz$ in the first term and again
integrating
by parts gives then
$$
(1-\a^3){\del\over\del\a}\vp_2\ =\ \a\vp_1 + 3 \a^2
\vp_2\ ,
$$
and hence
$$
{\del\over\del\a}\left({\vp_1\atop\vp_2}\right)\ =\
\pmatrix{ 0 & 1 \cr \coeff\a\oat & \coeff{3\a^2}\oat \cr}
\left({\vp_1\atop\vp_2}\right)\ .
$$
Upon eliminating $\vp_2$, this equation can be rewritten
$$
\Big[\oat\del^2_\a - 3\a^2\del_\a - a\Big]\,\vp_1\ =\ 0\
.\eqn\lintorus
$$
This example can easily be generalized to the series of
potentials,
$$
W\nn(x_A)\ =\ {1\over N}\sum_{i=1}^Nx_i^N\ -\
\a\prod_{i=1}^N x_i\ ,
\qquad\ \ N\geq3\ , \eqn\wnpots
$$
that describe \nex2 \LG\ theories with central charge
$c=3(N-2)$.
Consider the following periods, which are associated with
$(N-l-1,l-1)$-forms on
$V\nn$,
$$
\vp\nn_l\ =\ (l-1)!\!\int_\Gamma{\prod_{i=1}^N(x_i)^{l-
1}\over
(W\nn)^l}\,\om\ ,
\qquad\ \ l=1,\dots,N-1\ .
$$
We find an equation in  a ``Drinfeld-Sokolov'' form:
$$
\eqalignno{
&\dada{}.\vp\nn\ \cr &\ \ \ \ =\
\pmatrix{
0 & 1 & 0 & \cdots & 0 & 0  \cr
0 & 0 & 1 & \cdots & 0 & 0  \cr
0 & 0 & 0 & \cdots & 0 & 0  \cr
\vdots & \vdots & \vdots &  \ddots & \vdots & \vdots \cr
0 & 0 & 0 & \cdots & 0 & 1  \cr
\coeff{\a\bi N.1.}{\oam} & \coeff{\a^2\bi N.2.}{\oam} &
\coeff{\a^3\bi N.3.}{\oam} & \cdots &
\coeff{\a^{N-2}\bi N.N-2.}{\oam} &
\coeff{\a^{N-1}\bi N.N-1.}{\oam}    \cr   }\cdot \vp\nn\ \
,\cr
}
$$
where the coefficients are recursively defined:
$$
\bi N.l.\ =\ l\,\bi N-1.l.\, +\, \bi N-1.l-1.\ , \qquad {\rm
for}\ l=1,\dots,N-2\ ,
$$
with $\bi N.1.=1$ and $\bi N.N-1.=\shalf N(N-1)$.
The matrix equation yields
$$
\Big[\,\oam\,\dada N-1. - \big(\sum_{l=1}^{N-1}\bi N.l.\a^l
\,\dada l-1.\big)\, \Big]\,\vp_1\nn=\ 0\ .  \eqn\scalareq
$$
Note that due to the high degree of symmetry of the
perturbation,
the order of this equation is much less then the dimension
of the corresponding local ring.

Under the substitutions $\a^N\to z^{-1}$ and $\vp_1\to
z^{1/N}\vp_1$, this equation transforms into the following
generalized
hypergeometric differential equation \slat\ with regular,
singular
points at $z=0,1,\infty$ : \def\zdz{z{\del\over\del z}}
$$
\Big[\,\Big(\zdz\Big)^{N-1} -\ z\Big(\zdz+{1\over
N}\Big)\Big(\zdz+{2\over N}\Big)\dots \Big(\zdz+{N-1\over
N}\Big)\Big]\,\vp_1\nn=\ 0\ .  \eqn\hyper
$$
For \nex5, this is identical to the equation that was
discussed in \CandB:
\def\dzn#1{{\del^{#1}\over\del z^{#1}}}
$$
\Big[\,z^3(1-z)\dzn4+(6-8z)z^2\dzn3+(7-{72\over5}
z)z\dzn2+(1-{24\over5}z)\dzn{}-
{24\over625}\Big]\,\vp_1^{(5)} \ =\ 0\ .  \eqn\candeq
$$
Equation \hyper\ is solved \slat\ by
\def\Fn#1.#2.{\,{}_{#1}F_{#2}} \def\fn#1.{ {#1\over N}}
$$
\vp_1\nn\ =\ \Fn N-1.N-2.\Big[\,\fn1., \fn2., \dots,\fn N-1.
;\, 1, 1, \dots, 1;\,z\,\Big]\ ,
$$
where
$$
\eqalign{
\Fn A.B.\big[\,a_1, a_2,\dots, \a_A;\, b_1, b_2,\dots b_B;\,
&z\, \big] \cr
\equiv\ \ \ &{ \prod_{k=1}^B\Gamma(b_k) \over
\prod_{l=1}^A\Gamma(a_l) }  \,\sum_{n=0}^\infty\,{z^n\over
n!}\, {\prod_{l=1}^A\Gamma(a_l+n)\over
  \prod_{k=1}^B\Gamma(b_k+n)}   \cr}  \eqn\hyp
$$
is the generalized hypergeometric function. In direct
generalization
of the results of \CandB, a complete set of linear
independent solutions
to \hyper\ for $N>3$ is given by
$$
y_k\ =\ z^{-k/N}\Fn N-1.N-2.\Big[\,\fn k.,\fn k.,\dots,\fn
k.;
\overbrace{  \fn k+1.,\fn k+2.,\dots,\fn k+N-1.};\, z^{-1}
\,\Big] \eqn\sols
$$
($k=1,\dots,N-1$),
where the overbrace indicates that the entry with value
equal to one is to be omitted.

The reduction method that we have described for obtaining
the differential
equation \matdeq\ works far more generally than
for the class of potentials $W$ discussed above.
For example, the generalization to quasihomogenous spaces is
straightforward.
Moreover, one can apply these techniques
to marginal deformations of more general \LG\ models.
One does not have to restrict to \LG\ theories that have a
sigma-model
interpretation; one such generalization is obtained by
formally combining
theories with other ones so as to mimick the $c=3d$
situation.
Then one applies the results derived above and makes the
trivial observation
that so long as the marginal perturbations do not mix one
component theory with another, this tensoring of theories is
irrelevant
for the determination of the differential equation. Hence we
need not restrict to theories with $c=3d$. As an
illustration, consider
$$
W\ =\ x^3y + y^3 + z^3 - \a\,x^3z\ ,  \eqn\WU
$$
which describes a perturbation of an \nex2 theory with
$c=\coeff{11}3$ and
$\mu=14$. We find
$$
\Big[\,9\oat\dada 2.  -
24\a^2\dada{}. - 4\a\Big]\,\vp_1\ =\ 0\ .\eqn\deqU
$$
In general, for theories that have effectively one modulus,
the order of the differential equation will be two if
$3\leq c<6$, three if $6\leq c<9$, and so on.

In the next section, we will describe how the linear
equations \matdeq\ for
the periods are related to non-linear differential equations
that determine
the dependence of the flat coordinates on the moduli,
$\mu_i$.
In section 4, we will further generalize the method
to arbitrary marginal and relevant perturbations of generic
\LG\
potentials.

\chap{Non-Linear Equations, Duality and Monodromy Groups}

To obtain the flat or `special' coordinates on the moduli
space of a Calabi-Yau
manifold one expands the holomorphic $(3,0)$-form in a basis
of {\it integral}
cohomology \Strom\CandB.  That is, one introduces a
symplectically
diagonal basis, $\{\alpha_a, \beta^b\}$, of integral
cohomology and writes the
$(3,0)$-form as:
$$
\Omega ~=~ z^a~\alpha_a ~+~ {\cal G}_b ~\beta^b.
\eqn\OMdecomp
$$
(The periods $z^a$ and ${\cal G}_b$ are integral linear
combinations of the
entries of an appropriate row of the period matrix
$\Pi_\alpha^\beta$).
Since $\Omega$ is only defined up to multiplication by an
arbitrary function
$f(z^a)$, the quantities $z^a$ and ${\cal G}^b$ are only
defined projectively.
It was shown in \Strom\CandB\CandC\ that the $z^a$ define
good projective
coordinates on the moduli space, while the ${\cal G}_b$
satisfy ${\cal G}_b =
\partial_b(z^a {\cal G}_a)$.  It is thus the inhomogeneous
coordinates $\zeta^a =z^a/z^0$ that constitute the required
flat, or special coordinates, $t_i$.
The crucial ingredient that leads to the flat coordinates is
the choice of an {\it
integral} cohomology basis.  This, in a very strong sense,
means that we are
choosing a locally constant frame for the cohomology
fibration over the space of moduli.

 In the following, we will not restrict ourselves to $3$-
folds, but we will
consider projective coordinates $\zeta^a$ that are provided
by the expansion of
the the holomorphic $(d,0)$-form on a general `Calabi-Yau'
manifold.

In section 2 we saw how to derive a linear system of
equations \matdeq\ that is satisfied by the periods, $z^a$.
Obviously, if one multiplies $\Omega$ by $f(z^a)$ then one
will obtain a different set of equations, and thus \matdeq\
is not unique.
The appropriate {\it invariant} equation is a {\it non-
linear}
system of equations for $\zeta^a$ that can be derived from
the linear system  for $z^a$.

For example, the linear, second order equation
$$
{{d^2 z}\over {d \mu^2}} ~+~ p(\mu)\, {{d z}\over {d \mu}}
{}~+~ q(\mu)\,z
{}~=~ 0 \eqn\secordDE
$$
gives rise to the non-linear, Schwarzian differential
equation:
$$
\big\{ \zeta; \mu \big\} ~=~ 2\,I\ ,
  \qquad \ \ I\ \equiv\ q ~-~ \coeff14 p^2 ~-~ \shalf {{d
p}\over {d
\mu}}\eqn\SchwDE
$$
where $\zeta = z_1/z_2$ and $z_i \ $, $i= 1,2$ are the two
solutions of the
linear equation \secordDE. The Schwarzian differential
operator on the left-hand-side of this equation is defined
by:
$$
\big\{\zeta; \mu \big\} ~\equiv~ {{\zeta^{\prime \prime
\prime}}
\over {\zeta^\prime}} ~-~ {3 \over 2} \bigg( {{\zeta^{\prime
\prime}} \over {\zeta^\prime}} \bigg)^2\eqn\Schwdef
$$
and satisfies:
$$
 \big \{ y;x \big \} ~=~ - \big ( {{dy} \over {dx}} \big)^2
\big\{ x;y \big \} \eqn\propa
$$
$$
 \big \{ y;x \big \} ~=~  \big ( {{dz} \over {dx}} \big)^2
\Big [ \big\{ y;z \big \} - \big\{ x;z \big \} \Big ]
\eqn\propb
$$
$$
\big\{ y;x \big \} ~\equiv~ 0 \quad {\rm if~ and ~ only ~
if} \quad
y~=~ {{ax+b} \over {cx+d}} \, ,\eqn\propc
$$
for some $a,b,c,d \in \IC$ and $ad-bc \not = 0$.

The quantity $I$ in \SchwDE\ is often referred to as the
{\it invariant} of
\secordDE\ since it is unchanged if one replaces \secordDE\
by the linear
equation for $f(\mu) z(\mu)$ (where $f(\mu)$ is an arbitrary
function).

Recall \AGDJS\ that the `duality group' $\Gamma_W$ of the
super\-potential $W$
 consists of those transformations of the moduli that are
induced through
quasihomogeneous changes of the variables, $x^A$, that leave
the form of the
super\-potential unchanged up to an overall factor.  That
is, if $W_0 (x^B; \mu_i)$
is a quasi\-homogeneous potential then one seeks quasi\-
homogeneous changes
of variable $\hat x^A$ whose Jacobian, $det \big( {{\partial
\hat x^A} \over {\partial
x^B}} \big) $, is constant, or at worst a function,
$\Delta(\mu_i)$, of $\mu_i$ and
for which:
$$
W_0 (\hat x^A; \mu_i) ~=~  h(\mu_i)^{-1}~W_0 (x^A; \hat
\mu_i
(\mu_i))\, , \eqn\Wduality
$$
where $\hat \mu_i$ is some function of $\mu_i$.  If one
makes such a change of
variables in the period integrals of $\Omega$ then the
result is changed by an
overall factor of $h(\mu_i) \Delta (\mu_i)$.

 It follows that the linear system of differential equations
must be covariant with respect to the duality group of the
super\-potential.  That is, if $z(\mu_i)$ is a solution,
then so is
$ \Delta(\mu_i) h(\mu_i) z(\hat \mu_i (\mu_i))$.  The
corresponding system of
non-linear equations must be {\it invariant} with respect to
this duality group.  This duality covariance and invariance
can be very instructive in understanding the properties of
the linear and non-linear equations.
Turning this around, it allows in principle to determine
$\Gamma_W$
from the differential equations. In particular, given a
second order
equation \secordDE, one can often read off the solution of
the associated
Schwarzian equation in terms of triangle functions,
$s(\a,\be,\gamma)$.
The parameters $\a,\be$ and $\gamma$ then determine the
duality group of the
superpotential. For triangle functions, this group can be
thought of as being generated by reflections in the sides of
hyperbolic triangles (with angles $\pi\a,\pi\be$ and
$\pi\gamma$) that cover the upper half-plane.

It is thus of interest to understand the relationship
between the foregoing duality group $\Gamma_W$,
the monodromy group $\Gamma_M$ of the linear equations and
the `modular'
group $\Gamma$  of the surface.  The integral homology basis
undergoes an integral
symplectic transformation when it is transported around
singular points in the
moduli space of the manifold.  Consequently, the periods of
the differential forms undergo just such symplectic
transformations about these singular points.
This is directly reflected in the monodromy around regular
singular points of the
solutions of the differential equations.  The set of all
such monodromies will
generate a subgroup, $\Gamma_M$, of the `modular group',
$\Gamma$.  The set
of duality transformations $\Gamma_W$ of the superpotential
maps the surface
back to itself and will thus extend the group $\Gamma_M$ to
an even larger
subgroup of $\Gamma$.  In some cases this extension is all
of
$\Gamma$, and then the duality group of $W$ is
$\Gamma_W=\Gamma/\Gamma_M$.

We are uncertain as to the general validity
of this conclusion, but a simple illustration is provided by
the cubic torus,
\torusex. The non-linear system associated to \lintorus\ is
given by the Schwarzian
$$
\big\{\,\a;t\,\big\}\ =\ - {1\over2} {(8+\a^3)\over(1-
\a^3)^2}\,\a(\a')^2
 \ , \eqn\schwarz
$$
which is solved by the triangle function
$\at\!=\!s(\coeff12,\coeff13,\coeff13;J(t))$ \LLW\AGDJS.
The transformation properties of this function are well
known; in particular,
it is a modular form of $\Gamma(3)\equiv PSL(2,\ZZ_3)$, and
this group is the monodromy group $\Gamma_M$ of the
differential equation \lintorus. Both sides of \schwarz\ are
invariant under
$\Gamma=PSL(2,\ZZ)$, which is the full modular group of the
torus. The quotient, the tetrahedral group
$\Gamma/\Gamma(3)$, is precisely the
duality group $\Gamma_W$ of the superpotential \torusex,
\LLW\AGDJS.

As a further example of the non-linear systems of equations
that one can derive for the flat coordinates, we will
descibe in some detail the  a very particular family of $K3$
surfaces.  That is, we will consider the surface  defined by
the superpotential \wnpots\ for \nex4.  Equation \scalareq\
becomes:
$$
\Big[\,(1-\a^4)\dada 3.  - 6\a^3\dada2. -
7\a^2\dada{}. - \a\Big]\,\vp_1^{(4)}\ =\ 0 \ , \eqn\kthree
$$
Before discussing the solution of this system, it is
instructive to consider a general third order equation and
see how one passes to the associated non-linear system and
obtains the invariants.  Our discussion  will follow that of
\Forsyth.  Consider the generic third order equation:
$$
\vp''' + 3p(\a) \,\vp'' + 3 q(\a)\, \vp' + r(\a)\, \vp =0.
\eqn\genericde
$$
One starts by partially removing the freedom to multiply a
solution by an arbitrary function of $\alpha$.  This is done
by requiring the vanishing of
the coefficient of the second derivative, and is
accomplished by
substituting $ \vp \equiv y e^{ - \int p\, d\a}$.  The
differential equation then takes the form:
$$
y''' + 3 Q(\a)\, y' + R(\a)y\ =\ 0, \eqn\shortdeq
$$
where
$$
\eqalign{
Q &= q -p^{2} -p'
 \cr
R &= r-3pq + 2 p^{3} - p'' \ .\cr}\eqn\QR
$$
Let $y_1$, $y_2$ and $y_3$ be solutions of \shortdeq\ and
define $s$ and $t$ by
$$
s ~=~ \frac{y_{2}}{y_{1}}, \qquad t ~=~ \frac{y_{3}}{y_{1}}.
$$
Substituting $y_2 = s y_1$ and $y_3 = t y_1$ into \shortdeq,
and using the fact that $y_1$ is a solution, one obtains:
$$
\eqalign{ 3s^\prime y_1^{\prime \prime} ~+~ 3 s^{\prime
\prime} y_1^\prime
{}~+~ (3Q s^\prime ~+~ s^{\prime \prime \prime}) y_1 &~=~0 \cr
3t^\prime y_1^{\prime \prime} ~+~ 3 t^{\prime \prime}
y_1^\prime
{}~+~ (3Q t^\prime ~+~ t^{\prime \prime \prime}) y_1 &~=~0.}
\eqn\nonlina
$$
If one now differentiates these two equations again, and
eliminates $y_1^{\prime \prime \prime}$ using \shortdeq\ one
obtains two more equations  that are linear in $y_1$,
$y_1^\prime$ and $y_1^{\prime \prime}$.  These two
equations, along with \nonlina, provide four linear
equations for the three non-trivial, independent unknowns
$y_1$, $y_1^\prime$ and $y_1^{\prime \prime}$, and thus
there are two indepedent $3 \times 3$  determinants that
must vanish.  The vanishing of these determinants gives two
fourth order, non-linear equations for $s$ and $t$.
Conversely, given a solution to these non-linear equations,
one can eliminate $y_1^{\prime \prime}$ from the linear
system described above to obtain a simple linear, first
order equation for $y_1$, whose solution is:
$$
y_1 ~=~ \big(s^{\prime \prime} t^\prime ~-~ s^\prime
t^{\prime \prime} \big)^{-{1 \over 3}}.
$$
The other solutions are then obtained from $y_2 = s y_1$ and
$y_3 = t y_1$.

The actual non-linear system for $s$ and $t$ is fairly
unedifying, but we will give it here for the sake of
completeness.
Define the following variables:
$$
\eqalign{
u_{1} &= s''t' - s't'' \qquad   u_{2} = s^{(3)}\, t' -
s't^{(3)} \qquad
u_{3} = s^{(4)}\, t' - s't^{(4)} \cr
v_{1} &= s^{(3)}\, t'' - s''t^{(3)} \qquad v_{2} = s^{(4)}\,
t'' - s''t^{(4)},
\cr}\eqn\uuv
$$
where $s^{(i)} \equiv d^{i}s/d\a^{i}$, and
introduce the differential operators:
$$
\eqalign{
D_1(s,t;\alpha)\ & \equiv \ \frac{u_{3} - 2v_{1}}{u_{1}} -
\frac{4}{3} \left( \frac{u_{2}}{u_{1}} \right)^{2}  \cr
D_2(s,t;\alpha) \ & \equiv \ \frac{9 v_{2}}{u_{1}} - \frac{6
u_{2}(u_{3} + 4v_{1})}{u_{1}^{2}} + 8
\left( \frac{u_{2}}{u_{1}} \right)^{3}\ .
\cr}\eqn\DaDbdef
$$
The non-linear system may then be written:
$$
\eqalign{
D_1(s,t;\alpha)\ &=\ 3Q(\a) ~\equiv~ I, \cr D_2(s,t;\alpha)\
&=\ 27(\frac{\del Q}{\del\a} - R) ~\equiv~ J.}\eqn\yukkk
$$
The operators $D_1$ and $D_2$ are invariant under fractional
linear transformations:
$$
\eqalign{
D_1 \left( \frac{a_{2} + b_{2}s + c_{2} t}{a_{1}+ b_{1}s
+c_{1}t},\frac{a_{3}
+ b_{3}s + c_{3} t}{a_{1}+ b_{1}s
+c_{1}t};\a \right) &~=~ D_1(s,t;\a)  \cr
D_2 \left( \frac{a_{2} + b_{2}s + c_{2} t}{a_{1}+ b_{1}s
+c_{1}t},\frac{a_{3}
+ b_{3}s + c_{3} t}{a_{1}+ b_{1}s
+c_{1}t};\a\right) &~=~ D_2(s,t;\a)\ .
\cr}\eqn\DaDbinv
$$
The right-hand-sides of \yukkk\ define the quantities $I$
and $J$, which are called the {\it invariants} of the
system.

To solve \kthree\ one needs to use some more of the theory
of reduced differential equations of the form \shortdeq
\Forsyth \foot{A recent discussion of this subject may be
found in \DIZ.}.  The form of \shortdeq\ can be preserved by
a combined rescaling and reparametrization.  That is, one
introduces a new parameter $t$ and sets $y = \big( {{dt}
\over {d \alpha}} \big)^{-1} u$.  Under this transformation
the resulting differential equation has the form of
\shortdeq, but with:
$$
\eqalign{ \alpha &~\rightarrow~t \, , \qquad y ~\rightarrow~
u \, , \qquad Q \rightarrow~ \tilde Q ~\equiv~ \Big( {{dt}
\over {d \alpha}} \Big)^{-2} \Big(~Q ~-~ \frac{2}{3}
\big\{t; \alpha \big\} ~\Big) \cr
R &~\rightarrow~ \tilde R ~\equiv~ \Big( {{dt} \over {d
\alpha}} \Big)^{-3} \bigg[ \Big(~R ~-~ {d \over {d \alpha}}
\big\{t; \alpha \big\} ~\Big) ~-~ 3 \Big( {{d^2t} \over {d
\alpha^2}} \Big)~\Big( {{dt} \over {d \alpha}} \Big) \tilde
Q ~\bigg]\ .}
$$
It is interesting to note that $Q$ transforms precisely like
an  energy momentum tensor, and that
the combination $W_3 \equiv R - \frac{3}{2} \frac{dQ}{d
\alpha} $ transforms homogeneously, \ie:
$$
 W_3 \rightarrow~ \tilde W_3 ~\equiv~ \Big(~\tilde R ~-~
\frac{3}{2} \frac{d \tilde Q}{dt} ~\Big) ~=~ \Big( {{dt}
\over {d \alpha}} \Big)^{-3} W_3 \ .
$$
It is precisely one of the classical $W$-generators \DIZ.
One can fix the repara\-metriz\-ation invariance by
requiring that $\tilde Q = 0$, or
$$
\big\{t; \alpha \big\} ~=~ \frac{3}{2} ~Q \ .\eqn\newschw
$$
If one puts \kthree\ in the form \shortdeq\ one has:
$$
\eqalign{
Q\ &=\ \frac{\a^{2}}{3}\frac{(\a^{4} + 11)}{(1-\a^{4})^{2}}
\cr
R\ &=\  \a ~\frac{11+ 36\a^{4} + \a^{8}}{(1-\a^4)^{3}}\
,\cr}
$$
{}From this one finds an extra bonus: $W_3\equiv0$, or $R =
\frac{3}{2} Q'$.  This means that when  one passes to the
equation for $u(t)$, one obtains ${{d^3u} \over {dt^3}} =
0$, whose solutions are $1$, $t$ and $t^2$.  Therefore the
solutions to \kthree\ are:
$$
\vp ~=~ (1 ~-~ \alpha^4)^{-{1 \over 2}}~ \Big( {{dt} \over
{d \alpha}} \Big)^{-1} ~u (t)\ ; \qquad u(t) = 1\,, t\,,
t^2\, , \eqn\vpsoln
$$
where $t(\alpha)$ is the solution of \newschw:
$$
\big\{ t ; \a \big\}\ =\ {3\over2} \,Q\ \equiv\ {1\over2}
\a^{2} \left( \frac{\a^{4}+11}{(1-\a^{4})^{2}} \right)\
.\eqn\ktschw
$$
Finally, changing  variables $z=\alpha^{-4}$ in \ktschw\ one
obtains:
$$
\big\{t; z \big\} ~=~ \frac{1}{2} \frac{1}{z^2} ~+~
\frac{3}{8} \frac{1}{(z-1)^2} ~-~ \frac{13}{32}
\frac{1}{z(z-1)}.
$$
The solution of this equation is given by a triangle
function, $t(z)=s(0,\shalf,\coeff14;z)$,  which can, in
turn, be re-expressed as the ratio of two solutions to the
ordinary hypergeometric equation with parameters $\alpha
=\coeff18$, $\beta =\coeff38$, and $\gamma =1$. (The
solution can, of course, also be expressed in terms of
ratios of generalized
hypergeometric functions \sols.)
We remark that the structure of the monodromy group of
\kthree\ is
very similar to that of the quintic of \CandB, the
difference being that all appearances of $5$ in the formulae
of \CandB\ must be replaced by $4$.
This rule seems to hold for all $N$.

As we have seen, the modular dependence of the periods of
this family of $K3$ surfaces could have involved a non-
trivial $W_3$ invariant, but instead we found that $W_3$
vanished.  In this sense, the structure is determined merely
by the Virasoro algebra, that is, by the Schwarzian
differential equation \newschw.

It would be very interesting to discover to what extent the
higher dimensional surfaces defined by \wnpots\ might be
similarly reduced,
and to understand whether the appearance of such reduced
$W$-algebras has any deeper meaning. In particular, note
that the solutions of \kthree\ are algebraically related
(inspection of \vpsoln\ shows that $\vp_1 \vp_3 = \vp_2^2$),
and as we have seen this is a consequence of the vanishing
of $W_3$.
It turns out that for the quintic, \ie\ for \candeq, one
also has $W_3\equiv0$
but $W_4\not=0$, and it is known that the $\cG_b$ in
\OMdecomp\ are homogenous
functions of the $z^a$. Thus it appears that the vanishing
of $W$-generators
is closely connected to algebraic relations between the
solutions.
 We hope to discuss these issues elsewhere.

\penalty-1000
\chap{Flat Coordinates for Generic Perturbations}

We now wish to generalize the methods of section 2 to
marginal and relevent perturbations of arbitrary
topological \LG\ field theories\foot
{Our discussion will follow, and
extend, that of \BVar. Flat coordinates in generic
topological
matter theories  have also recently been discussed in
\CeVa.}.
 The basic problem is that general \nex2 superconformal
theories have no obvious analogue of {\it integral}
cohomology.
As discussed in the previous section, it is this that leads
one to flat coordinates for `Calabi-Yau' spaces.

For general topological matter models, one can make a
general ansatz for ${\cal
F}$, or for $W$, in terms of the flat coordinates, and then
evolve algebraic and
differential equations from consistency conditions of the
topological matter
models \DVV\EVNW\ZMa.  In particular one requires that
the $C_{ijk}$ be given by \CFreln\ and that they satisfy
\Ccons.
However, solving the system \Ccons\ is extremely laborious,
except in the simplest cases.

Thus we like to obtain differential equations that determine
the flat coordinates more directly from the superpotential.
Let $W_0(\xa)$ be a quasihomogenous superpotential and
$W$$(\xa;\si)$ be a para\-me\-tri\-za\-tion of a general,
versal deformation
of $W_0$ by elements $\phi_\a$ of the chiral ring.
The problem is to determine the relationship between the
general coordinates $\si$, and the flat coordinates $t_i$.
Once the parametrization of $W$ in terms of the $t_i$ is
known, the free energy $\cF$ and
all correlation functions can easily be computed.

We will regard $W(\xa;\si)$ as a quasihomogenous function of
$\xa$ and
$\si$, and thus the coupling constants $\si$ can be assigned
dimensions.
(We will adopt the convention that both $W_0$ and $W$ have
dimension
equal to one.)
Below we will actually consider only marginal and relevant
perturbations,
whose corresponding coupling constants will have vanishing
or positive
dimensions. This will lead to the major simplification that
all quantities
will have polynomial dependence on the coupling constants
with positive
dimension, and the only non-polynomial behavior will be via
the marginal
parameters.

The coupling constant associated to the constant term in
$W(\xa;\si)$
(\ie, the unique coupling constant of dimension one) will
play a
distinguished, important role and it will be denoted by
$\so$. The
remaining coupling constants $\s_2, \s_3,\dots$ will be
denoted
generically by $\sp$. We will take
$$
{\widetilde W}(\xa,\sp)\ =\ W(\xa,\si) - \so \eqn\eqone
$$
as independent of $\so$.

Let $\pa,\ \a=1,\dots,\mu$, be any (polynomial) basis for
the chiral ring
and consider integrals of the form
$$
\u\a\l.\ =\ \mol\lp1.\G\lp1.\int_\gamma\!\w\pa.\lp1.\dx  \ ,
\eqn\eqtwo
$$
where the integral is taken over any compact homology
cycle\foot
{There are $\mu$ independent possible choices, but the
choice is
not important for the moment. The reader might find it
helpful to consider
the one-variable case, in which $\gamma$ is some loop around
some subset of the zeros of $W$.} $\gamma$ in the set
$\{x\in \IC^n:
W(x,s)\not=0\}$.
The gamma function and the factor of $\mol\lp1.$ are
introduced for later
convenience.
These integrals are related to the periods of differential
forms
on the level surfaces of $W$, \Arn.
They satisfy some important recurrence relations
\Saito\BVar:
$$
\eqalignno{
\del^k_\so\,\u\a\l.\ &=\ \u\a\lp k.\ ,\qquad k\in\ZZ
\phantom{\sum_k^k}
&\eqnalign{\eqthree}
\cr
\del_\si\,\u\a\l.\ &=\ \ \ \ \sum_{k=-1}^\infty\Bia
k.\,\u\be\l-k.
&\eqnalign{\eqfour}
\cr
\so\, \u\a\l.\ &=\ \,-\sum_{k=0}^\infty\Aa k-2.\,\u\be\l-k.
&\eqnalign{\eqfive}
 \cr
}
$$
These recurrence relations are derived by the same procedure
as that employed
in section 2. Equation \eqthree\
is a trivial consequence of differentiation under the
integral. Equation \eqfour\ is
also obtained by differentiating under the integral, but in
this second instance the
numerator of the integrand is a polynomial $(\del_\si W)\pa$
which might need to
be reduced.
That is, by definition of the local ring of $W$ this
polynomial
may always be rewritten in the following form:
$$
(\del_\si W)\pa\ \equiv\ {C_{i \a}}^\be\phi_\be(\xa;\sp) +
      q^{(0)A}_{i\ \a}(\xa;\sp)\,{\del W\over
\del\xa}(\xa;\sp)
\eqn\eqsix
$$
for some polynomial $q^{(0)A}_{i\ \a}$. One now integrates
by parts\foot
{That is, one uses the fact that
$0\equiv\int\ddx\left(\w V^A.\lp1.\right)\dx$ for any vector
$V^A$.}
to obtain
$$
\del_\si\,\u\a\l.\ =\ {C_{i \a}}^\be\u\be\lp1. +
       \mol\lp1.\G\lp1.\int\w(\ddx q^{(0)A}_{i\
\a}).\lp1.\dx\ .
$$
Once again one decomposes the numerator
$$
\ddx q^{(0)A}_{i\ \a}\ =\ \Bia 0.\,\phi_\be(\xa;\sp) +
      q^{(1)A}_{i\ \a}(\xa;\sp)\,{\del W\over
\del\xa}(\xa;\sp)\ ,
$$
and integrates by parts. In this manner one may recursively
compute
$\Bia k.$, $k=-1,0,1,\dots$. Note that after every
integration by parts
the polynomial degree (in $\xa$) of the successive terms
$(\ddx q^{(k)A}_{i\ \a})$ decreases by one unit, and hence
this
procedure must terminate after a finite number of steps.
Also note that
${C_{i \a}}^\be\equiv\Bia -1.$ are essentially the structure
constants of the
local ring.

Finally, equation \eqfive\ is obtained by taking $\so$
inside the integral,
and rewriting it as $\so\equiv W(\xa,\si)-{\widetilde
W}(\xa,\sp)$.
The factor of $W$ is cancelled immediately, while
${\widetilde W}(\xa,\sp)\pa$ is simplified by the identical,
recursive
reduction procedure described above.

It is very convenient to make a ``Fourier transformation''
in the
$\so$ variable. Specifically, it replaces $f(\so)$ by
$\int_{-\infty}^0e^{\so/z}f(\so)\,d\so$. This has the effect
of sending
$\del_\so\to z^{-1}$ and $\so\to-z^2{d\over dz}$.
Let $\u\a\l.(z;\sp)$ denote the transform of $\u\a\l.(\s)$.
Then
equations \eqfour\ and \eqfive\ may be rewritten as a linear
system:
$$
\eqalignno{
\left(\del_\si - \sum_{k=-1}^\infty z^k \Bia
k.\right)\,\u\be\l.(z;\sp)
  \ &=\ 0
  &\eqnalign{\eqseven}
  \cr
\left(\del_z - \sum_{k=-2}^\infty z^k \Aa
k.\right)\,\u\be\l.(z;\sp)
  \ &=\ 0
  &\eqnalign{\eqeight}
  \cr
}
$$
Observe that the $\u\be\l.(z;\sp)$ are, by definition,
covariant constant
sections of a flat vector bundle whose connections are
defined
by $\cB_i^{(k)}$ and $\cA^{(k)}$.

To get more insight into why we make this construction,
suppose that
$\cB_i^{(k)}\equiv0$ for $k\geq1$ and define
$\cD_i=\del_\si-B_i^{(0)},\
{C_{i a}}^\be=\Bia -1.$. Then the flatness of the
connection, or
integrability of \eqseven\ implies
$$
\big[\cD_i - z^{-1} C_i\,,\cD_j-z^{-1} C_j\big]\ =\ 0\
,\eqn\eqnine
$$
and separating out different orders in $z$, one gets the
zero curvature
equations\foot{Such equations have also been discussed in
\Strom\CeVa.}
$$
\big[\cD_i\,,\cD_j\big]\ \equiv\ \cD_{[i}\,C_{j]}\ \equiv\
\big[ C_i\,, C_j\big]\ \ \equiv\ 0\ . \eqn\eqten
$$
Hence the connection $\cB_i^{(0)}$ is flat, the structure
constants
${C_{i a}}^{\be}$ commute and are covariantly constant. If
we now
arrange that the basis $\{\phi_\a\}$ of the chiral ring is,
in fact,
given by $\{{\del W\over\del\si}\}$, and let
$$
{\Gamma_{i  j}}^{k} =\ \cB_{i j}^{(0)\,  k}\ ,
$$
then one finds that  ${\Gamma_{i j}}^{ k}=
{\Gamma_{j  i}}^{ k}$ and equation
\eqten\ implies that $\Gamma$ is the flat coordinate
connection we seek.
We thus have solved the consistency conditions \CFreln\ and
\Ccons.
The remainder of this section will essentially reduce the
general
problem to the foregoing simpler situation.

Because the connection defined by $\cA$ and $\cB$ is flat,
one already knows that one can find a gauge transformation
that will
trivialize it. More precisely, because there might be non-
trivial
monodromy, one can find a matrix $M$ such that \foot{These
matrices are
analytic in $\sp$, but not in $z$, hence the form of the
equation.}
$$
\eqalign{
\cA\ &=\ (\del_zM)M^{-1} + M \big({\tilde A(\sp)\over
z}\big) M^{-1}\cr
\cB_i\ &=\ (\del_{\si}M)M^{-1}\ . \cr
}\eqn\eqelev
$$
Thus we can gauge away all of $\cB$ and almost all of $\cA$.
The problem is that the matrix $M$ will in general involve
all
powers in $z$ and $1/z$. Hence $M$ will define a basis
change involving
$\u\a\lp k.$ for all $k\in\ZZ$. To control this, and indeed
to preserve
quasihomogeneity, we want to restrict ourselves to changes
of basis
that are upper triangular, that is, $\u\a\lp k.$ is only
modified
by addition of polynomials in $\sp$ and $\u\a\lp l.$ for
$l\leq k$. This means that the change of basis must be
analytic at $z=0$.
Now suppose that we can decompose $M$ of equation \eqelev\
in the following
manner:
$$
M\ =\ g_0(z;\sp)\,g_\infty(\coeff1{z};\sp)\ ,\eqn\eqtwelve
$$
where $g_0$ is analytic at $z=0$ and $g_\infty$ is analytic
at $z=\infty$.
Now let $\cA'\equiv g_0^{-1}\cA g_0-g_0^{-1}(\del_zg_0)$ and
$\cB_i'\equiv g_0^{-1}\cB_i g_0-g_0^{-1}(\del_zg_0)$. Then
it is elementary to
see that
$$
\eqalign{
\cA'\ &=\ \Big[(\del_zg_\infty)g_\infty^{-1} +
g_\infty\big({A_0(\sp)\over
z}\big)g_\infty^{-1}\Big] \cr
\cB_i'\ &=\ (\del_{\si}g_\infty)g_\infty^{-1}\ .\cr
}\eqn\eqthirt
$$
Moreover, by modifying $g_0$ by multiplying by a suitable
matrix,
$h(\sp)$, one can further gauge away the $z$-independent
term in
$\cB'$. Thus, provided that we can make the split in
\eqtwelve\ there is
a $z$-analytic gauge choice that has
$\cA^{(k)}\equiv\cB_i^{(k)}\equiv0,
k\geq0$.

The problem of finding the splitting \eqtwelve\ is called a
Riemann-Hilbert
problem, and is generically \LCrn, but not always, solvable.
Its solution is intimately connected with solving integrable
models (see, for
example, \FaT).
We have, in fact, a variational Riemann-Hilbert problem in
that our
matrices have parameters $\sp$. This makes the problem much
easier to
address and it will be discussed further in the appendix. In
particular,
we show in the appendix that $g_0=I +\cO(\sp)$, where $I$ is
the identity matrix,
and we will also show that
$g_0$ is analytic in $\sp$ and preserves quasihomogeneity
(\ie\ the elements
of the new basis have a well-defined scaling dimension).

To get flat coordinates, we need to make the restriction
to marginal or relevant perturbations (of dimension less
than or equal to
one), which means $dim(\si)\geq0$.
Let $v_\a(z;\sp)$ be the basis in which
$\cA^{(k)}\equiv\cB^{(k)}\equiv0$
for $k\geq 0$. Observe that if we restrict to $\pa$ of
dimension strictly less
than one, then the corresponding $\u\a\l.(\sp)$ have
dimensions strictly
less than $(\sum \omega_A)-\l$ (where $\omega_A$ is the
weight of
$\xa$). However, because the basis change has the form:
$g_0={\bf 1}+\cO(\sp)$,
is analytic in $z$ and $\sp$ and preserves quasihomogeneity,
it follows that the
$v_\a$
of dimension strictly less than $(\sum \omega_A)-\l$  are
analytic,
quasihomogenous combinations of $\sp$ and the
$\u\a\l.(\sp)$.  In particular,
such $v_\alpha$ do not involve any $u_\alpha^{(\lambda +
k)}(\sp)$ for $k \not =
0$). Furthermore,
the $v_\a$ of dimension equal to $(\sum \omega_A)-\l$ must
be analytic,
quasihomogenous combinations of $\sp$, the $\u\a\l.$, and
$$
u_0'\ \equiv\ \u0\l-1.\ \equiv\ \mol\l.\G\l.\int\w1.\l.\dx\
.\eqn\fourt
$$
One of the $v_\a$ of dimension equal to $(\sum \omega_A)-\l$
must be (a dimensionless multiple of) $\u0\l-1.= z\u0\l.$,
while the rest of the
$v_\a$
must start with a $\u\a\l.$ term for which $\pa$ is a
marginal operator.
Let $\tilde s$ denote the (dimensionless) marginal
parameters and
let $v_\a'$ and $f(\tilde s)$ be such that $\{v_\a',
f(\tilde s)u_0' \}$  forms a set of
linearly independent $v_\a^{(\l)}$ of
dimension less than or equal to $(\sum \omega_A)-\l$.

It follows from the foregoing that there is a
quasihomogenous, analytic,
invertible matrix $e_\a^{\ \ j}(\sp)$ and a set of functions
$q_j(\tilde\s)$ such that
$$
{\del\over\del\s_j}u_0'\ =\ e_j^{\ \ \a}\,v_\a' -
q_j(\tilde\s)\,u_0'
\eqn\eqfift
$$
and $q_j\equiv0$ if $dim(\s_j)>0$. Next observe that
$$
\eqalign{
{\del^2\over\del\si\del\s_j}u_0' \ &=\ \big( \del_ie_j^{\ \
\a} \, \big)\,v_\a'
+e_j^{\ \ \a}(\del_iv_\a')
        - q_j(\del_i u_0') - (\del_iq_j)u_0'   \cr
&=\ \big[\del_ie_j^{\ \ \a}-q_je_i^{\ \ \a}\big]v_\a' +
z^{-1}e_j^{\ \ \a}{C_{i a}}^\be v_\be'
 + \big[q_iq_j-\del_iq_j\big]u_0'\ .\cr}
$$
By linear independence of the $v_\a'$ and $u_0'$ we have
$\del_{[i}\,q_{j]}\equiv0$ and {\it hence}
$q_j\equiv\del_jq$ for
some function $q(\tilde\s)$.

The foregoing combines to give us the following simple
result. There is
a `universal' function $q(\tilde\s)$ of all the marginal
(dimension
zero) parameters such that the integral
$$
u_0\ =\ \mol\l.\G\l.\int\w q(\tilde\s).\l.\dx   \eqn\eqsixt
$$
satisfies the following equation\foot
{Remember that $W$ is only perturbed by marginal and
relevant operators, \ie,
$0\leq dim(\s_j)\leq1$.}
$$
{\del^2\over\del\si\del\s_j}u_0\ =\ {C_{i
j}}^{\a}\,\u\a\lp1. +
    {\Gamma_{i j}}^{ k}\,\Big({\del\over\del\s_k}\Big) u_0\
. \eqn\eqsevent
$$
Thus, the function $q(\tilde\s)$ is determined by requiring
that
\eqsevent\ contains no term proportional to $u_0$ itself.
The ${C_{i j}}^{\a}$ are just the structure constants of the
chiral ring and ${\Gamma_{i j}}^{k}$ is the Gauss-Manin
connection.
Flat coordinates are determined by simply requiring that
$\Gamma\equiv0$
on the r.h.s.\ of \eqsevent.

In practice one takes the $\si$ to be the flat coordinates
$t_i$,
and considers a perturbation of the form
\def\aa{{i}}
$$
W(x;t) \ =\  W_0(x) + \sum \mu_\aa(t)\, m_\aa(x)\ ,
\eqn\pertw
$$
where $m_\aa(x)$ are monomials in the local ring (with
degree less or equal
than one), and the \LG\ couplings, $\mu_\aa(t)$, are unknown
functions to be
determined.
We note that it is elementary to explicitly write down the
constraints
implied by \eqsevent\ since this only involves
differentiating under the
integral and integrating by parts, just like the reduction
procedure
described in section 2. The constraints take first the form
of linear differential
equations for the $\mu_\aa$. One determines $q(\tilde t)$
in terms of the $\mu_\aa(t)$ by requiring that the $u_0$
piece in
\eqsevent\ vanishes.
Substituting for $q(\tilde t)$ then
turns the linear system into the associated non-linear
system
(\eg, into a Schwarzian differential equation) that
determines
the $\mu_\aa(t)$.

The function $q(\tilde t)$ appears to be playing the r\^ole
of a conformal
rescaling of the vielbein. In particular we note that for
the examples we
computed, the function $q(\tilde t)^{-2}$ is precisely the
conformal
factor that takes the Grothendieck metric of \CeVa\ to the
flat
metric.

For conformal theories with $c > 3$, there are chiral
primary fields of dimension larger than one.  With the
restriction that we have made on the perturbed
superpotential, we cannot write these irrelevant chiral
primaries as ${{\partial W} \over {\partial s_i}}$ for some
$s_i$.  Thus it might appear that these irrelevant chiral
primaries play no role in determining the form of the
equations that we derive from the procedure described above.
This is not so.  It is important to remember to pass first
to the basis for {\it all} the chiral primaries in which one
has $\cB_i^{(k)} = 0$ for $k \ge 0$, and then one must use
this basis in calculating the separate terms in expressions
like those on the right-hand-side of \eqsevent.

Finally we note that
equations \eqsevent\ and \eqten\ can be recast in the
familiar form
$$
\tilde \cD_i\,\vp\ =\ 0 \qquad \ \ {\rm and}\ \ \ \
\big[\tilde \cD_i\,,\tilde \cD_j\big] \ =\ 0\ ,\eqn\matrixeq
$$
where $\tilde \cD_i=\del_{\si}+\Gamma_i+C_i\del_1$ and
$\vp_i=\del_iu_0$.
The first equation is a generalization of the matrix
differential
equation \matdeq\ we discussed in section 2.

\chap{Examples Revisited}

It is instructive to reconsider first the torus example
\torusex\ of section 2,
but now with an additional, relevant perturbation:
$$
W\ =\ \coeff13(x^3+y^3+z^3) - \a(t)\,xyz-s\,\b1(t)xy -
\shalf s^2\b2(t)z
   -\coeff16s^3\b3(t)\ .
 \eqn\Wexample
$$
 Here, $t$ is a dimensionless, flat coordinate (the modular
parameter
 of a torus), and $s$ is a parameter of dimension $1/3$.
 The dependence of the \LG\ coupling constants on the
relevant perturbation
 parameter $s$ is already fixed by its dimension, so we will
have
 to determine only the dependence on the modular parameter.
 The two-parameter perturbation is certainly not the most
general one
 (which was considered previously in \EVNW),
 but the extension is obvious. The specific perturbation we
chose is
 however the most general one consistent with the
$\ZZ_3\times\ZZ_3$
 symmetry generated by
 \def\om{\omega}
 $(x,y,z,s)\to(\om x, \om^2y,z,s)$ and $(x,y,z,s)\to(\om
x,\om y,\om z,
 \om s)$.

 Let $u_0\equiv\mol\l.\G\l.\int\!\w q(t).\l.\dx$. We want to
solve
 for $\at$ and $ \b i(t)$ by requiring the connection
$\Gamma$
 in \eqsevent\ to be flat; this corresponds to the vanishing
of terms
 proportional to $u_\a^{(\l)}$ in this equation. In
particular, we obtain
 $$
 \eqalign{
 {\del^2u_0\over\del t^2}\ &=\ \left({q''\over q}\right)u_0
  \ +\ \mol\lp1.\,\G\lp1.\intw1\ \ \times \cr
  \ \ \ \ &\Big\{\sBig\}
  \Big[2{q'\over q}+{\a''\over \a'}\Big]\a'q\,xyz +
  s\Big[2{q'\over q}+{\b1''\over \b1'}\Big]\b1'q\,xy  \cr
  &\qquad\qquad\ \ \ \ \ +\shalf s^2\Big[2{q'\over
q}+{\b2''\over \b2'}\Big]\b2'q\,z +
  \coeff16 s^3\Big[2{q'\over q}+{\b3''\over \b3'}\Big]\b3'q
   \sBig\{\Big\} \cr
  +\ &\ \ \mol\lp2.\,\G\lp2.\intw2 \ \ \times \cr
  \!\!\!&\Big\{\sBig\}\,(\a')^2q\,x^2y^2z^2 + 2s\a'\b1'q\,
       x^2y^2z + s^2(\b1')^2q\,x^2y^2 + s^2\a'\b2'q\,xyz^2 +
   s^3[\b1'\b2'\cr &\!\!+\coeff13\a'\b3']q\,xyz + \coeff14
s^4(\b2')^2q\,z^2
   + \coeff13s^4\b1'\b3'q\,xy + \coeff16 s^5\b2'\b3'q\,z +
   \coeff1{36}s^6(\b3')^2q
   \sBig\{\Big\}\ .
  \cr}\eqn\seconder
  $$
 We could obtain equations for $q,\a, \b i$ by considering
all the different powers
of
 $s$ in this equation, but it is easier to just concentrate
on the $s=0$ pieces. We
can thus use \xyzexp\
  and subsequently $z^3\attac{s=0}= (z \del_zW+\a
xyz)\attac{s=0}$ to integrate
 $\int \w x^2y^2z^2.\lp2.$ by parts to reduce its degree.
The vanishing
 of the connection $\Gamma$ corresponds to the vanishing of
the terms
proportional to $\w1.\lp1.$, \ie
 $$
 \mol\lp1.\G\lp1.\intw1\Big[2{q'\over q}+{\a''\over \a'}
 +{3 \a^2\a'\over \oat}  \Big]\a' q\,xyz\ \equiv\ 0\ .
 $$
 This determines $q(t)$
 $$
 q(t)\ =\ \Big({1-\at^3\over\a'(t)}\Big)^{1/2} \ .\eqn\qdet
 $$
 Integrating \seconder\ (with $s=0$) by parts, substituting
\qdet\ and
requiring the vanishing of all terms that are proportional
to $u_0$,
we then indeed obtain directly the Schwarzian differential
equation \schwarz\ for $\at$,
$$
\big\{\a;t\big\}\ =\ -
{1\over2}{(8+\a^3)\over\oat^2}\a(\a')^2\ ,
$$
which is associated to the linear equation \lintorus.
However, by using the methods derived in the foregoing
section, we can
now also solve for the couplings $\b i(t)$ of the relevant
perturbations.
To obtain $\b1(t)$, it is easiest to consider the $s=0$
piece of
${\del^2u_0\over\del s\del t}$; by integrating by parts, we
find the
condition
$$
 \w1.\lp1.\Big[2{q'\over q}+{\b1'\over \b1}+{2 \a^2\a'\over
 \oat} \Big]\b1 q\,xy\ \equiv\ 0\ ,
$$
and this gives
$
\b1(t) = A (\a(t)')^{1/2}(1-\at^3)^{1/6}
$
(where $A$ is an integration constant).
Similarly, from the $s$ independent piece of
${\del^2u_0\over\del s^2}$
we obtain
$$
\w1.\lp1.\Big[\b2+{\b1^2\a\over\oat}\Big]q\,z\ \equiv\ 0\ ,
$$
which yields
$
\b2(t) = - A^2\at\at'(1-\at^3)^{-2/3}
$.
Finally, for $\b3$ we consider the piece linear in $s$ of
${\del^2u_0\over\del s^2}$, and use the identity
$$
\eqalign{
x^2y^2 \ &=\ {1\over\oat}\Big\{\a xz\del_xW + x^2\del_yW +
\a^2xy\del_zW\cr
&\ \ \qquad +2s\a\b1\,xyz + s\b1x\del_xW + s^2\b1^2 + \shalf
s^2\a^2\b2 xy
 \Big\} \cr}
$$
Partial integration gives
$$
\int\w1.\lp1. s q\Big[\b3+{\b1^3\over\oat}\Big]\ \equiv\ 0$$
and thus determines $\b 3(t)$.
These results coincide with the expressions derived in
\EVNW.
One can also check that for the choice of $\a, \b i$ give
above, the connection $\Gamma$ is completely flattened.

To illustrate that our method may be used for theories with
arbitrary
central charges,
reconsider the potential \WU\ with additional, relevant
perturbations:
$$
W\ =\ x^3y + y^3 + z^3 - \at\,x^3z - s_1\be_1(t)\,z -
s_2\be_2(t)\,x^2\ ,
$$
We find for the $\w1.\lp1.$ piece in the $s_i=0$ part of
${\del^2u_0\over\del t^2}$
$$
 \w1.\lp1.\Big[2{q'\over q}+{\a''\over \a'}-{8\over3}
 {\a^2\a'\over(1+\a^3)}
  \Big]\a' q\,x^3z\  \ ,
$$
and this determines
$$
q(t)\ =\ \a'(t)^{-1/2}\big[1+\at^3\big]^{4/9}\ .
$$
The $u_0$ piece in \eqsevent\ then vanishes if $\a$
satisfies the
differential equation
$$
\big\{\a;t\big\}\ =\  {40\over9(1+\a^3)^2}\a(\a')^2\ .
$$
This is precisely the Schwarzian form of the linear equation
\deqU.
Moreover, $\be_1$ may be obtained from the $s_i=0$ piece in
${\del^2u_0\over\del t\del s_1}$:
$\be_1(t)=\a'(t)^{1/2}[1+\at^3]^{-1/3}$.
Similarly, we find $\be_2(t)=\a'(t)^{1/2}[1+\at^3]^{-1/9}$.
Hence, we
can compute the following term of the free energy:
$$
\cF(\si,t)\ =\ \shalf s_1 {s_2}^2 \a'(t)^{1/2}
[1+\at^3]^{1/9}\ +\  \dots\ .
$$
It is clear that we could compute the other terms of $\cF$
in a similar way.

\vskip 1.truecm
\centerline{\bf Acknowledgements}

The work of W.L.\ was supported by DOE contract DE-
AC0381ER40050 and that of D.-J.S. by NSF grant PHY 90-21139.
N.W.\ is supported in part by funds provided by the DOE
under grant No. DE-FG03-84ER40168 and also by a fellowship
from the Alfred P. Sloan foundation.

D.-J.S.\ thanks O.\ Alvarez and A. Giveon for discussions.
N.W. would like to thank D. Morrison
and S. Katz for telling him about the reduction procedure
described in section 2 of this paper.  He would also like to
thank K. Uhlenbeck for very helpful
discussions on the Riemann-Hilbert problem.  D.-J.S. and
N.W. are grateful to the MSRI in Berkeley and the organizers
of the Workshop
on Mirror Symmetry for creating a stimulating environment in
which some of the ideas in this paper were developed.
W.L. and N.W. would also like to thank the Aspen Center for
Physics
for providing a excellent environment that was very
conducive to the further progress of this research.

While this
manuscript was in preparation we received the preprint
\Ferr\ in which
similar results to those of section 2 are derived.

\nobreak\vskip 1.truecm
\nobreak\appendix

Our purpose here is to show why one can find a matrix
$g_0(z,\sp)$ that is analytic in $z$ and $\sp$ in the
neighbourhood of $z=0$, $~\sp = 0$ and which will gauge the
potentials ${\cal
A}^{(k)}$ and ${\cal B}^{(k)}$ to zero for $k \ge 0$. As was
shown in section 3, it
suffices to show that the matrix $M(z;\sp)$ defined in
\eqelev\  can be factorized
as in \eqtwelve.  The Birkhoff decomposition theorem \LCrn
\GDB\ implies that
any matrix $M(z; \sp)$ can be decomposed according to:
$$
M\ =\ g_0(z;\sp) ~ \Lambda(z;\sp) ~ g_\infty(\coeff1z;\sp)\
,\eqn\appone
$$
where $g_0$ is analytic at $z=0$,  $g_\infty$ is analytic at
$z=\infty$ and
$\Lambda(z;\sp)$ is a diagonal matrix whose entries are
integral powers \foot{
These integers are related to the Chern class of the
relevant vector bundle over
$S^2$.} of $z$.  We need to show that $\Lambda = I$, where
$I$ is the identity
matrix.  More simply, it suffices to show that all the
integral powers of $z$
in $\Lambda$ are, in fact, zero and hence $\Lambda$ is $z$
independent and
can thus be absorbed into $g_0$ or $g_\infty$.  Since
integers can only be
continuous functions of $\sp$ by being constant, we can
establish the desired
result in a region about $\sp = 0$ by simply showing that it
is true at $\sp = 0$.

Let $\phi_\alpha(x) =
\phi_\alpha(x,\sp = 0)$, and recall that $W_0(x)  \equiv
\widetilde
W(x,\sp = 0)$.  By assumption $W_0(x)$ and $\phi_\alpha(x)$
are
quasihomogeneous of weight $1$ and of some weight
$\l_\alpha$
respectively.  It follows that $W_0 \equiv \sum_A \omega_A
x^A{\del W_0
\over\del\xa}$ and hence
$$
\eqalign{ W_0(x) ~ \phi_\alpha(x) ~\equiv~ & \sum_A \omega_A
x^A ~
{{\partial W_0} \over {\partial x^A}} ~ \phi_\alpha (x)\cr
{}~=~ &
\sum_A \bigg[ {{\partial} \over {\partial x^A}} \big(
\omega_A x^A
W_0 \phi_\alpha \big) ~-~ \omega_A  W_0 \phi_\alpha ~-~
\omega_A x^A
W_0 {{\partial \phi_\alpha} \over {\partial x^A}} \bigg] \cr
{}~=~ &
\sum_A  {{\partial} \over {\partial x^A}} \big( \omega_A x^A
W_0
\phi_\alpha \big) ~-~ \Big[ \big(\sum_A \omega_A \big) +
\l_\alpha
\Big] W_0 \phi_\alpha .}
$$
Therefore
$W_0(x)~ \phi_\alpha (x)$ is a total derivative at
$\sp = 0$.  It follows that ${\cal A}^{(k)}(\sp) \equiv 0$
for $k \ge 0$.
Now take
$$ g_0(z,\sp) ~=~ I ~+~  \sum_j \sum_{k = 0}^{\infty} z^k
s_j^\prime {\cal B}_j^{(k)} (\sp = 0) ~+~ O((\sp)^2)$$
where $I$ is the identity matrix.  This yields the desired
gauge for
all values of $z$ but with $\sp = 0$.  As described above,
the Birkhoff theorem
then guarantees that it can be done in a region about $\sp =
0$.  Also note that
$g_0(z,\sp = 0) ~=~ I $.

To see that the solution can be made quasihomogeneously,
consider the
differential equation that needs to be solved:
$$\eqalign{ g_0^{-1} {\cal A} g_0 ~-~ g_0^{-1} \partial_z
g_0 ~=~ &
z^{-2} P_{-2} (\sp) ~+~ z^{-1} P_{-1}(\sp) \cr g_0^{-1}
{\cal B}_j g_0 ~-~ g_0^{-1} \partial_{s_j} g_0 ~=~ & z^{-2}
Q_{j}
(\sp)}$$
where $P_{-1}$, $P_{-2}$ and $Q_j$ are unknowns.  This means
that we
must solve:
$$\partial_{s_j} g_0 ~=~ g_0 \Big[ g_0^{-1} {\cal B}_j g_0
\Big]_+ \eqn\appone$$
where $[~~~]_+$ means: take only the non-negative powers of
$z$ in a
power series expansion about $z=0$.  The fact that the
system
\appone\ is integrable follows from the general observations
above. (It could
probably be proved more directly.) If $g_0$ is known to
$n$-th
order in $\sp$, then \appone\ determines the $(n+1)$-st
order term.
Thus one can evolve a power series in $\sp$. (The
convergence of this power
series is guaranteed by the Cauchy-Kowalewski theorem.)   It
is elementary
to see that quasihomogeneity is preserved order by order.
Finally, it is of some
interest to know over what size of patch one can find the
required gauge.
Generically one finds \KUh\ that this gauge choice can be
made in a large
Schubert cell of the underlying Lie group.  That is, one
expects
that $g_0(z;\sp)$ will become singular when one runs into a
Weyl point of the
underlying Lie group.

\vskip 3.truecm

 \pagegoal=7.truein\parskip=0.truein
   \immediate\closeout\referencewrite
   \referenceopenfalse
   \line{\hfil{\fourteenpoint\rm References}\hfil}
   \vskip\headskip
   \input referenc.texauxil

\end